\documentclass[11pt]{article}

\usepackage{amsmath,amssymb}
\usepackage{graphicx}
\usepackage{stmaryrd}
\usepackage{overpic}
\usepackage{hyperref}

\makeatletter

\renewcommand{\section}{\@startsection{section}{1}{\z@}
                                   {-3.5ex \@plus -1ex \@minus -.2ex}
                                   {2.3ex \@plus .2ex}
                                   {\normalfont\large\bfseries}}
\renewcommand{\subsection}{\@startsection{subsection}{2}{\z@}
                                   {-3.25ex\@plus -1ex \@minus -.2ex}
                                   {1.5ex \@plus .2ex}
                                   {\normalfont\normalsize\bfseries}}
\renewcommand{\subsubsection}{\@startsection{subsubsection}{3}{\z@}
                                   {-3.25ex\@plus -1ex \@minus -.2ex}
                                   {1.5ex \@plus .2ex}
                                   {\normalfont\normalsize\bfseries}}
\renewcommand{\paragraph}{\@startsection{paragraph}{4}{\z@}
                                   {3.25ex \@plus1ex \@minus.2ex}
                                   {-1em}
                                   {\normalfont\normalsize\bfseries}}

\makeatother

\newcommand{\be}{\begin{equation}}
\newcommand{\ee}{\end{equation}}
\newcommand{\bea}{\begin{eqnarray}}
\newcommand{\eea}{\end{eqnarray}}
\newcommand{\ba}{\begin{array}}
\newcommand{\ea}{\end{array}}

\newcommand{\id}{\hbox{1\kern-.27em l}}
\newcommand{\ZZ}{\mathbb{Z}}

\newcommand{\RR}{\mathbb{R}}
\newcommand{\lb}{\langle}
\newcommand{\rb}{\rangle}

\newcommand{\sss}{\scriptscriptstyle}

\newcommand{\ga}{\gamma}
\newcommand{\Ga}{\Gamma}
\newcommand{\bet}{\beta}

\newcommand{\de}{\delta}

\newcommand{\ep}{\epsilon}
\newcommand{\si}{\sigma}
\newcommand{\la}{\lambda}

\newcommand{\De}{\Delta}
\newcommand{\La}{\Lambda}

\newcommand{\cN}{\mathcal{N}}
\newcommand{\cO}{\mathcal{O}}
\newcommand{\cM}{\mathcal{M}}
\newcommand{\cA}{\mathcal{A}}
\newcommand{\cL}{\mathcal{L}}
\newcommand{\cD}{\mathcal{D}}
\newcommand{\cH}{\mathcal{H}}
\newcommand{\cP}{\mathcal{P}}

\newcommand{\Tr}{{\rm Tr}}

\newcommand{\vac}{|0 \rangle}

\newcommand{\vacg}{|0 \rangle_g}

\newcommand{\al}{\alpha}
\newcommand{\ald}{\alpha^{\dagger}}
\newcommand{\lap}{\lambda_+}

\newcommand{\lam}{\lambda_-}

\newcommand{\lapm}{\lambda_{\pm}}
\newcommand{\lapmbar}{\overline{\lambda}_{\pm}}
\newcommand{\lamp}{\lambda_{\mp}}
\newcommand{\lampbar}{\overline{\lambda}_{\mp}}

\newcommand{\SU}{\mathrm{SU}}

\newcommand{\SO}{\mathrm{SO}}
\newcommand{\SL}{\mathrm{SL}}

\newcommand{\Spin}{\mathrm{Spin}}

\newcommand{\g}{\mathfrak{g}}
\newcommand{\su}{\mathfrak{su}}
\renewcommand{\u}{\mathfrak{u}}

\newcommand{\iso}{\cong}

\begin{document}
\pagenumbering{arabic}
\thispagestyle{empty}

\begin{center}
\vspace*{35mm}
{\LARGE \bf \sf Finite energy shifts in $\SU(n)$ supersymmetric\\[2mm] Yang-Mills theory on $T^3\times\RR$ at weak coupling}

\vspace*{10mm}
{\large Fredrik Ohlsson}

\vspace*{5mm}
\textit{Department of Fundamental Physics\\
Chalmers University of Technology\\
S-412 96 G\"oteborg, Sweden}\\[3mm]
{\tt fredrik.ohlsson@chalmers.se}     

\vspace*{35mm}{\bf Abstract:} 
\end{center}
\vspace*{5mm}
\noindent We consider a semi-classical treatment, in the regime of weak gauge coupling, of supersymmetric Yang-Mills theory in a space-time of the form $T^3\times\RR$ with $\SU(n)/\ZZ_n$ gauge group and a non-trivial gauge bundle. More specifically, we consider the theories obtained as power series expansions around a certain class of normalizable vacua of the classical theory, corresponding to isolated points in the moduli space of flat connections, and the perturbative corrections to the free energy eigenstates and eigenvalues in the weakly interacting theory. The perturbation theory construction of the interacting Hilbert space is complicated by the divergence of the norm of the interacting states. Consequently, the free and interacting Hilbert furnish unitarily inequivalent representation of the algebra of creation and annihilation operators of the quantum theory. We discuss a consistent redefinition of the Hilbert space norm to obtain the interacting Hilbert space and the properties of the interacting representation. In particular, we consider the lowest non-vanishing corrections to the free energy spectrum and discuss the crucial importance of supersymmetry for these corrections to be finite.

\newpage



\setcounter{equation}{0}
\section{Introduction}
\label{sec:Introduction}
In $3+1$ dimensions there are three distinct classes of pure supersymmetric Yang-Mills theories, labelled by the amount of supersymmetry, $\cN=1,2$ and $4$ respectively, that they possess. A SYM theory is completely specified by the choice of $\cN$, a compact gauge group with simply connected cover $G$ and the gauge coupling constant $g$. The generic field content is a gauge field $A^{\mu}$, scalar fields $\Phi$ and spinors $X$, all transforming in the adjoint representation of the Lie algebra $\g$ of $G$. In the case of extended supersymmetry the scalars and spinors also furnish non-trivial representations of the R-symmetry. We will consider these theories in a space-time of the form $T^3 \times \RR$, where $\RR$ denotes the time and it is assumed that the spatial three-torus is $T^3 = \RR^3/\ZZ^3$. The torus preserves all supersymmetries and translational invariance while breaking the continuous Lorentz symmetries, and in the case of $\cN=4$ the conformal symmetry, present in a Minkowski space-time. Due to its periodicity the torus also introduces a natural infrared cut-off, because the spatial momentum vectors have a finite shortest length, removing any IR divergences of the theories. In this space-time geometry we will be concerned with the weak coupling energy spectrum around certain normalizable vacua of the SYM theories. As we will discuss in more detail later, the compactness of the spatial $T^3$ implies that we indeed expect non-trivial energy corrections since scattering states cannot be separated to asymptotically free regions.  

Throughout, we will work in a Hamiltonian formalism in temporal gauge, $A^0 \equiv 0$, with gauge group $G_{adj}=G/C_G$, where $C_G$ is the centre subgroup of $G$. In this setting, the wave-functions of vacuum states in the weak coupling limit are localized on the moduli space $\cM$ of gauge inequivalent flat connections, which generically consists of several disconnected components. The moduli space is parameterized by the holonomies around the generators of the fundamental group of the torus, which define an almost commuting triple of elements in $G$. The components of $\cM$ are characterized by the topology of the corresponding gauge bundle and the rank of the subgroup of $G_{adj}$ left unbroken by the almost commuting triple. Points in $\cM$ where the gauge group is completely broken correspond to normalizable vacua of the classical theory and it is therefore possible to perform a power series expansion to obtain a weak coupling description of the theory at such points. These matters will be discussed in more detail in sections two and three.

The free limit, i.e.~where $g\to 0$, of the theory at rank zero points of $\cM$  was treated in \cite{Lindman:2008}, where the spectrum was computed for arbitrary choice of $G$. Special emphasis was put on the case of $\cN=4$ but, as mentioned, the results obtained easily generalize to $\cN=1$ and $\cN=2$. In the present paper we attempt to continue the analysis in \cite{Lindman:2008} by considering the lowest order perturbative corrections to this spectrum in the special case where $G = \SU(n)$. We will describe the $3+1$ dimensional theory as the dimensional reduction, discussed in section four, of $\cN=1$ SYM theory in higher dimensions. This point of view provides a suitable framework for studying the weak coupling spectrum and allows the generalization of the results of the present paper from the $\cN=4$ case to $\cN=1$ and $\cN=2$.

After a brief review of the free theory in the higher dimensional $\cN=1$ context in section five, we proceed with a perturbative treatment of the interacting theory in section six. We first consider the Hilbert space of the interacting theory to linear order in $g$ and derive formal expressions for the energy corrections to quadratic order using perturbation theory in $g$. The states produced by standard perturbation theory, expressed as linear combinations of states in the free Hilbert space, turn out to have infinite norm and do consequently not belong to the free Hilbert space. However, the Hilbert space norm can be consistently redefined to render the norms finite, a procedure which amounts to defining a new Hilbert space, inequivalent to that of the free theory. The two distinct Hilbert spaces correspond to unitarily inequivalent representations of the algebra of the creation and annihilation operators of the free theory. We discuss in detail the construction of the interacting Hilbert space, and the corresponding representation, from the states produced by perturbation theory. As we will see, the standard perturbation theory formulas for the energy shifts are valid. However, it is not obvious that they give finite results. We investigate the energy corrections to $\cO(g^2)$ and discuss how supersymmetry entails a non-trivial cancellation of the UV divergences, rendering the corrections finite. 

The finiteness of the energy corrections is certainly expected in the $\cN=4$ theory, which is known to be finite to all orders. The $\cN=1$ and $\cN=2$ theories are also consistent quantum theories and their interacting energy spectra are also expected to be finite after the theories are renormalized. The present consideration of the energy eigenvalues, however, does not require the implementation of any renormalization scheme to yield finite corrections. Furthermore, the construction of the interacting Hilbert space contains, as mentioned above, several subtleties that must be addressed to consistently define the Hilbert space of the interacting theory using perturbation theory. It should finally be stressed that the explicit perturbative corrections to the energy are only numerically accessible, and beyond the scope of the present paper.

\setcounter{equation}{0}
\section{Vacuum states in Yang-Mills theory on $T^3$}
\label{sec:VacuumStatesInYangMillsTheoryOnT3}
The choice of temporal gauge, $A^0 \equiv 0$, and the fact that the space-time we are considering is a direct product $T^3\times \RR$, allows us to consider the fields of the SYM theory as sections of fibre bundles over the spatial part $T^3$ with an additional time-dependence. We consider the adjoint form of the gauge group $G_{adj}=G/C_G$, where $C_G \subset G$ is the centre subgroup of $G$. In this context, the gauge field $A^i$, where $i=1,2,3$ denote the spatial dimensions, is the connection of a principal $G_{adj}$-bundle $P$ over the base space $T^3$. Through the adjoint action of $G$ on its Lie algebra $\g$ we can construct the associated vector bundle
\be
\label{eqn:AssociatedVectorBundle}
E = \mathrm{ad}(P) = P \times_{\mathrm{ad}} \g \,.
\ee
The scalar and spinor fields appearing in the various SYM theories are then sections of the bundles $E$ and $E\,\otimes\,S$ respectively, where $S$ is the unique spinor bundle over the full space-time compatible with supersymmetry.

\subsection{Bundle topology}
Since the base manifold is three-dimensional, the isomorphism class, or topological class, of the gauge bundle $P$ is completely determined by its discrete magnetic abelian 't Hooft flux (or Stiefel-Whitney class)
\be
\label{eqn:MagnetictHooftFlux}
\hat{m} \in H^2(T^3,C_G)\,.
\ee
This can be understood by considering the first few homotopy groups of compact simple Lie groups $G_{adj}$. Both $\pi_0(G_{adj})$ and $\pi_2(G_{adj})$ are trivial while $\pi_1(G_{adj}) \iso C_G$ due to the fact that $G$ is the simply connected covering group of $G_{adj}$. Consequently, the only way non-trivial bundle topology can arise is if the transition functions of the bundle wrap non-trivial one-cycles in the gauge group $G_{adj}$. The 't Hooft flux measures the obstruction to lifting the principal $G_{adj}$-bundle to a principal $G$-bundle over $T^3$ and completely specifies the topology of the bundle, thus determining $P$ up to isomorphisms.

The 't Hooft flux $\hat{m}$ is in turn completely specified by its restrictions $m_{ij} \in H^2(T^2,C_G) \iso C_G$ to the two-tori in the $i$ and $j$ directions, implying that we can express $\hat{m}$ as a triple of elements in the centre subgroup
\be
\label{eqn:MagnetictHooftFluxTriple}
\hat{m} = (m_{23},m_{31},m_{12}) \in C_G^3\,.
\ee  
This triple transforms as a vector under the mapping class group $\SL(3,\ZZ)$ of the torus and if the centre subgroup is cyclic, which is indeed the case for the gauge groups $G=\SU(n)$ we will be considering, it is possible to choose coordinates on $T^3$ such that $\hat{m} = (\id,\id,m)$ and the topological class of the bundle is encoded by the single element $m \in C_G$. Since $E$ and $E\,\otimes\,S$ inherit their topology from $P$ through the associated bundle construction, $m$ determines the bundle isomorphism classes completely.

\subsection{Vacuum states and flat connections}
A low-energy state of the Yang-Mills theory is characterized by the vanishing of the magnetic contribution, proportional to $\Tr(F^{ij}F^{ij})$, to the energy density. This implies that such states are supported on the moduli space $\cM$ of flat connections, i.e.~connections with $F^{ij}=0$. Furthermore, the electric energy contribution is proportional to $\Tr(F^{0i}F^{0i})$, implying that vacuum states are locally constant on $\cM$ since the momentum conjugate to $A^i$ is $F^{0i}$. 

The moduli space $\cM$ is parameterized, modulo simultaneous conjugation, by the holonomies
\be
\label{eqn:Holonomies}
\hat{U}_i = \cP \left ( \exp \, i \int_{\ga_i} A \right ) \,,
\ee
where $\ga_i$, $i=1,2,3$, are the three homotopically inequivalent generators of the fundamental group $\pi_1(T^3)$ of the torus. Since the curve $\ga_i\ga_j\ga_i^{-1}\ga_j^{-1}$ is contractible the holonomies constitute a triple $(\hat{U}_1,\hat{U}_2,\hat{U}_3)$ of mutually commuting elements in $G_{adj}$. The lifting $(U_1,U_2,U_3)$ of the commuting triple to the universal cover $G$ will generally not commute, due to the obstruction to lifting the bundle $P$ to a principal $G$-bundle. However, it satisfies the relation
\be
\label{eqn:AlmostCommutingtriple}
m_{ij} = U_i U_j U_i^{-1} U_j^{-1}\,,
\ee
where $m_{ij}$ are the components of the 't Hooft flux $\hat{m}$. The elements $U_i$ are referred to as an almost commuting triple. Since the centre $C_G$ acts trivially on $\g$, it is possible to simultaneously diagonalize the adjoint action of the $U_i$ on the Lie algebra $\g$ by choosing a basis $T_z$ satisfying
\be
\label{eqn:DiagonalBasis}
U_i^{-1} T_z U_i = z_i T_z\,.
\ee
The eigenvalues $z_i$ are complex roots of unity by virtue of the finite order of the $U_i$, due to the cyclic structure of $\pi_1(T^3)$, and form an eigenvalue vector $\vec{z}$. The action of the mapping class group $\SL(3,\ZZ)$ on the holonomies induces an action on the eigenvalue vector by element multiplication. We note that the adjoint action of $\hat{U}_i$ is also diagonalized by the $T_z$ and that the eigenvalues $z_i$ are not affected by the lifting procedure.

Flat connections with distinct topology, described by $m$, constitute disjoint subspaces $\cM(m) \subset \cM$ which are generally disconnected. The holonomies may break the gauge group to a subgroup $H \subset G_{adj}$, which is the centralizer or commutant of the triple $\hat{U}_i$, and the rank of $H$, here denoted $r_a$, is constant on each component of $\cM(m)$
\be
\label{eqn:ModuliSpace}
\cM(m) = \bigcup_a \cM_{r_a}\,.
\ee
Generically, $H$ is abelian but at certain subspaces $\cM^H$ it may be enhanced with non-abelian terms, so that its Lie algebra is
\be
\mathfrak{h} = \mathfrak{s} \oplus \u(1)^r\,,
\ee
where $\mathfrak{s}$ is semi-simple of rank $r_{\mathfrak{s}}$ and $r_a = r_{\mathfrak{s}} + r$.

In particular, the gauge group may be completely broken, $r_a=0$, to a finite group $H$, in which case the almost commuting triple has no eigenvalue vector $\vec{z} = (1,1,1)$. The corresponding components $\cM_0$ are isolated points (zero-dimensional subspaces) in $\cM(m)$. In fact, there is a one-to-one correspondence between isolated points in $\cM$ and rank zero triples. Almost commuting triples $U_i$ that completely break the gauge group are therefore referred to as rank zero or isolated triples. The conjugacy classes of isolated triples have been extensively studied in mathematics \cite{Borel:1953,Borel:1961,Borel:1999} and in terms of the application in the context of supersymmetric Yang-Mills theory described above \cite{Witten:1998, Keurentjes:1998, Kac:1999, Keurentjes:1999a, Keurentjes:1999b, Witten:2000, Henningson:2007a, Henningson:2007b, Henningson:2008}. 

In addition to a flat connection and vanishing conjugate momentum to the gauge field $A^i$, a vacuum state of SYM theory is characterized by covariantly constant scalar and spinor fields. Modes associated with broken generators become massive, contributing a finite energy to the Hamiltonian density, which means that they must vanish in zero energy field configurations. In the following section we will restrict considerations to connections that completely break the gauge group. Such vacua are characterized by the vanishing of all scalar and spinorial modes.

\setcounter{equation}{0}
\section{The weak coupling expansion}
\label{sec:TheWeakCouplingExpansion}
In this paper we are concerned with the limit where the gauge coupling is weak, i.e.~where the coupling constant is small $g \ll 1$. In this limit a semi-classical treatment of the quantum theory can be obtained by expanding the fields in powers of $g$ around any zero-energy field configuration and quantizing the fluctuations. In order to obtain a finite theory in the expansion, we must demand that the vacuum state used as a background is normalizable. Given a gauge group $G$, the first question to address is thus which vacua of the classical theory, if any, are normalizable and suitable for a weak coupling expansion.

To answer the question we consider the low energy effective field theories localized at the subspace $\cM^H \subset \cM$, corresponding to unbroken gauge group $H$. When $\mathfrak{h}$ contains abelian terms the vacuum states discussed in the previous section are not normalizable, since each $\u(1)$ term corresponds to a flat direction in the phase space of the theory. The solution to the equation of motions in these directions are therefore plane waves in field space that cannot be normalized. We must consequently restrict $H$ to be semi-simple or finite.

At subspaces $\cM^H$ with $H$ semi-simple, the quantized low energy effective theory is described by supersymmetric quantum mechanics with gauge group $H$ and $4,8$ or $16$ supercharges, corresponding to $\cN=1,2$ and $4$ respectively \cite{Witten:2000}. These theories have been extensively studied in the context of $D$-brane physics and it is conjectured that the only cases which admits normalizable zero energy configurations are the theories with $16$ supercharges \cite{Witten:1995}. For the case $G=\SU(n)$ we will be considering, the possible semisimple parts of the unbroken subgroups $H$ are direct products of $\SU$-factors. The supersymmetric quantum mechanics for such factors supposedly has a single bound states at threshold which is normalizable \cite{Witten:1995,Sen:1996}. However, the problem of constructing the corresponding wave-functions, or even proving the existence of these states, has proven very difficult and remains an open one. In order to establish the validity of a weak coupling expansion and perform explicit calculation we require a thorough understanding of the vacuum state of the classical theory. Such an understanding is not yet obtained for the bound states at threshold, on account of their elusive nature, and we will therefore restrict all further considerations to the final remaining case where $H$ is finite.

In contrast to $H$ semi-simple the case when the connection breaks the gauge group to a finite subgroup is trivial, due to the vanishing of all scalar and spinor field modes and the isolation of such a connection in $\cM$ discussed in the previous section. Thus, specifying a flat connection, or equivalently an almost commuting triple, that completely breaks the gauge group is equivalent to specifying a normalizable vacuum state of the classical theory. The structure of the moduli space $\cM$ can be determined using the results obtained in \cite{Borel:1999} where, in particular, a complete classification of almost commuting rank zero triples for compact, connected and simply connected $G$ was given. 

Given an isolated flat connection $\cA^{\mu}$ we can now perform the weak coupling expansion, mentioned above, of the fields in powers of $g$. Such an expansion yields
\be
\label{eqn:WeakCouplingExpansion}
\left \{ \ba{lll}
A^{\mu} & = & \cA^{\mu} + g a^{\mu} \\
\Phi & = & g \phi \\
X & = & g \chi
\ea \right. \,
\ee
since all scalar and spinor modes vanish in the corresponding zero energy field configuration. This expression can be considered exact, including all orders of the expansion in the perturbations $a^{\mu}$, $\phi$ and $\chi$. The covariant derivative with respect to the flat connection $\cA^i$ is denoted $\cD_i$, while the covariant derivative with respect to the full gauge field $A^i$ is denoted $D_i$. The choice of temporal gauge implies $a^0 = 0$, and to fix the remaining redundant gauge degree of freedom we choose to impose the Coulomb gauge condition, $\cD_i a^i = 0$, on the fluctuations in the connection. 

\subsection{The covariant derivative and momenta on $T^3$}
It is convenient at this point to return to the vector bundle $E = \mathrm{ad}(P)$ considered in the previous section. Sections of this bundle are Lie algebra-valued functions of space-time, and in particular we are interested in the space $\Ga(E)$ of $L^2$ sections of $E$, where the norm is taken with respect to the sesquilinear inner product
\be
\label{eqn:InnerProduct}
(\al,\bet) = \int_{T^3} d^3x\,\Tr(\overline{\al}\bet)\,.
\ee
The self-adjoint covariant derivatives with respect to the background connection constitute endomorphisms of this space 
\be
\label{eqn:SelfAdjointCovariantDerivative}
i \cD_i \, : \, \Ga(E) \to \Ga(E)\,,
\ee
and commute by flatness of $\cA$, implying that they can be simultaneously diagonalized. Simultaneous eigensections $u_p(x)$, satisfying
\be
\label{eqn:EigenvalueEquationCovariantDerivative}
i\cD_i u_p = 2 \pi p_i u_p \,,
\ee
were constructed in \cite{Lindman:2008} through parallel transport of the Lie algebra basis elements $T_z$ by the holonomy
\be
\label{eqn:Holonomy}
\tilde{g}(x) = \cP \left ( \exp \, i \int_{0}^x \cA \right )
\ee
according to
\be
\label{eqn:Eigenfunctions}
u_p(x) = \tilde{g}(x)^{-1} T_z \tilde{g}(x) e^{-log z_i x^i}\,.
\ee
The relation between the eigenvalues $z_i$ and the momentum eigenvalues $p_i$ is simply given by
\be
\label{eqn:EigenvalueRelation}
p_i = \frac{1}{2\pi} \mathrm{Arg}z_i + k_i\,,
\ee
where $k_i \in \ZZ$ and $\mathrm{Arg}z_i$ is the principal argument of $z_i$. Thus, the non-abelian part of $\cD_i$ shifts the momenta admitted by the torus by a rational number in the range $[0,1)$. A characteristic feature of the isolated flat connections we are considering here is that at least one of the components $p_i$ will always receive a non-zero shift, since there are no $\vec{z}=(1,1,1)$ eigenvalue vectors. In \cite{Lindman:2008} the eigenvalues $z_i$ were computed for all isolated flat connections in the classification of \cite{Borel:1999}.

Returning to the properties of $u_p$ we find that, with a suitable scaling of the generators of $\g$, they constitute an orthonormal basis of $\Ga(E)$ satisfying
\be
\label{eqn:EigenfunctionProperties}
\ba{ccc}
\overline{u}_p & = & u_{-p}\\
(u_p,u_{p'}) & = & \delta_{p,p'}\\
\sum_p u_p^a(x) \overline{u}_p^b(x') & = & \delta^{ab} \delta^{(3)}(x-x')
\ea \,,
\ee
where $a,b$ are indices taking values in the Lie algebra $\g$. This basis can be used to perform a Fourier expansion, the details of which will be considered in the next section. The Fourier coefficients are then promoted to creation and annihilation operators in the quantum theory to obtain the momentum space formulation of the SYM theory.

In computing the spectrum of the theory we will need to compute the Lie bracket of the momentum eigenfunctions. From the construction of $u_p$ we find that
\be
\label{eqn:LieBracketMomentumEigenfunctions}
[u_p(x),u_{p'}(x)] = C_{p,p'} u_{p+p'}(x) \,,
\ee
where the coefficients $C_{p,p'}$ are defined by the structure constants of $\g$ according to
\be
\label{eqn:CPPprime}
[T_z,T_{z'}] = C_{p,p'} T_{z \cdot z'}\,,
\ee
where $[\cdot,\cdot]$ should not be confused with the commutator of operators in the quantum theory introduced below. In $z \cdot z'$ the $\cdot$ denotes multiplication in each component of the eigenvalue vectors. This action corresponds to the addition of the momentum eigenvalues $p_i$ and $p'_i$. We see that $C_{p,p'}$ is antisymmetric in $p$ and $p'$, and depends only on the basis vectors $T_z$ and $T_{z'}$ associated to the two momenta. In particular, this implies that $C_{p,p'}\neq 0$ only for $p$ and $p'$ corresponding to distinct generators $T_z$ and $T_{z'}$. Furthermore, in the basis used above for the Lie algebra $\g$ the coefficients $C_{p,p'}$ are purely imaginary.

In the general case the set of $\vec{z}$ is not closed under the action of $\cdot$, which corresponds to empty subspaces in the $\ZZ^3_r$ gradations of the Lie algebra $\g$, where $r$ is an integer, defined by the triple $(U_1,U_2,U_3)$ \cite{Lindman:2008}. If $z \cdot z'$ labels such an empty subspace the generators $T_z$ and $T_{z'}$ commute and the corresponding coefficients $C_{p,p'}$ vanish.

It should be noted that the spatial momentum operator $P_i$ acting on the Hilbert space of the SYM theory does not receive any corrections in $g$ in the weak coupling expansion (\ref{eqn:WeakCouplingExpansion}) while the Hamiltonian $H$ does receive such corrections. The reason for the qualitative difference between these two quantities is the fact that the manifold $T^3\times \RR$ we are considering has spatial periodicity but not temporal periodicity. Consequently, the eigenvalues of the generators of spatial translation must belong to the lattice reciprocal to $T^3$, and are therefore protected from corrections arising from continuous deformations of the theory. The eigenvalues of the Hamiltonian on the other hand, are not similarly restricted and can indeed receive corrections. Consequently, the momenta $p_i$ admitted by the torus are found as the eigenvalues of the $i\cD_i$ operator even for finite coupling strength. The eigensections $u_p$ thus form a suitable basis of $\Ga(E)$ in the Hamiltonian formalism also at finite $g$.

\subsection{The $G = \SU(n)$ case with non-trivial topology}
So far, the discussion is valid for arbitrary choice of gauge group $G$ and isolated flat connection $\cA$. We will now restrict our attention to a special choice of gauge group, namely $G=\SU(n)$. The corresponding Lie algebra $\su(n)$ has the appealing feature of non-degenerate momentum eigenvalues\footnote{For all other choices of $G$ the isolated points in the moduli space of flat connections correspond to degenerate momentum eigenvalues \cite{Lindman:2008}.}, implying that the eigenfunctions $u_p$ are completely characterized by the triple $p_i$.

The centre subgroup is $C_{\SU(n)} \iso \ZZ_n \iso \{v\id\,|\,v^n=1\}$ which is cyclic, allowing us to characterize the bundle topology by a single element $m \in C_{\SU(n)}$. Furthermore, we know from the results of \cite{Borel:1999} that the moduli space $\cM$ for $\SU(n)$ gauge groups only contains isolated points for certain non-trivial isomorphism classes of bundles, corresponding to non-trivial bundle topology, namely those where $m$ is a generator $c$ of $C_{\SU(n)}$. There are consequently $\varphi(n)$ classes of bundles admitting isolated flat connections and the corresponding moduli spaces are
\be
\label{eqn:ModuliSpaceSUn}
\cM_{\SU(n)}(m=c) = \bigcup_{i=1}^{n} \cM_0^{\sss(i)}\,,
\ee
where $\varphi(n)$ is the Euler $\varphi$-function counting the number of integers less than or equal to $n$ which are coprime to $n$. We recall that the subscript of the components of $\cM_{\SU(n)}(m=c)$ is the rank $r_a$ of the unbroken gauge group, implying that the gauge group is completely broken on $\cM_0^{\sss(i)}$. Here, we have also added an additional superscript $(i)$ counting the different components with coinciding values of $r_a$. Note that the subspaces $\cM_{\SU(n)}(m=c)$ contain no components in addition to the isolated points. 

Corresponding to the $n$ rank zero components for a particular choice of $c$, associated to a particular $n$:th root of unity $v$, are the $n$ almost commuting triples in $\SU(n)$ with
\be
\label{eqn:IsolatedTriplesSUn}
\ba{ccc}
U_1 = \left(\ba{cc}0&\id_{n-1}\\(-1)^{n-1}&0\ea\right) &,&
U_2 = a_n \cdot \mbox{diag}(1,v,\ldots,v^{n-1}) \,,
\ea
\ee
where
\be
a_n = \left\{ \ba{cll}
1&,&n = 2p+1\\
v^{1/2}&,&n=2p \ea \right. \,,
\ee
and $U_3$ any element of $C_{\SU(n)}$. Indeed these elements are almost commuting and satisfy $m_{12}=c$ and $m_{i3}=\id$ as required. The triples, unique up to conjugation, correspond to the same triple in $G_{adj} = \SU(n)/\ZZ_n$ since the centre of $\SU(n)$ acts by multiplication on the third component $U_3$ \cite{Borel:1999} and the magnetic 't Hooft flux completely characterizes the components of $\cM$ for a gauge group on the adjoint form. The different choices of the generator $c$ do, however, correspond to inequivalent conjugacy classes of rank zero triples, implying that there are $\varphi(n)$ distinct triples that completely break the gauge group and thus define normalizable vacuum states of the classical theory suitable as background field configurations for the weak coupling expansion.

The spectrum of the $i\cD_i$ operator for all the $\varphi(n)$ rank zero triples is
\be
\label{eqn:SpectrumSUn}
\left(\ba{c}p_1\\p_2\\p_3\ea\right) \in \left\{ \left. \left(\ba{c}\frac{q_1}{n}+\ZZ\\\frac{q_2}{n}+\ZZ\\\ZZ\ea\right) \right|
\ba{l}
q_1,q_2 \in \{1,\ldots,n\}\\
(q_1,q_2) \neq (n,n)
\ea
\right\} \,,
\ee
which is non-degenerate as mentioned above. The fact that the spectrum is identical for all the available triples implies that the weak coupling expansion will not depend on which of the $\varphi(n)$ vacuum states we choose as the background configuration.

The remainder of this paper is devoted to the study of the weak coupling expansion of the SYM theory for the case $G=\SU(n)$ and arbitrary choice of compatible background field configuration. The general approach and discussion will be applicable to any choice of gauge group $G$, but the perturbation theory analysis in section six has to be suitably modified to accommodate degenerate momentum eigenvalues.

\setcounter{equation}{0}
\section{The minimally supersymmetric perspective}
\label{sec:TheMinimallySupersymmetricPerspective}
Supersymmetric Yang-Mills theory in $3+1$ space-time dimensions with $\cN$ supercharges can be viewed as the dimensional reduction of $\cN=1$ SYM in $d+1$ dimensions \cite{Brink:1977} which is defined by the Lagrangian density 
\be
\label{eqn:MinimalLagrangian}
\cL = \frac{1}{g^2} \Tr \left\{ -\frac{1}{4} F_{MN}F^{MN} + \frac{i}{2} \overline{\psi} \Ga^M D_M \psi \right\} \,.
\ee
The $\cN=2$ theory is obtained for $d=5$ and the $\cN=4$ theory for $d=9$. The vector index takes values $M=0,1,\ldots,d$ where the spatial part will be denoted $I=1,\ldots,d$ to distinguish it from the three dimensions of the spatial torus, denoted $i=1,2,3$. In (\ref{eqn:MinimalLagrangian}) $\psi$ is a Majorana spinor in $3+1$ dimensions, a Weyl spinor in $5+1$ dimensions and a Majorana-Weyl spinor in $9+1$, yielding matching numbers of bosonic and fermionic degrees of freedom in each case. As we will see, this higher dimensional perspective on the four dimensional SYM theory is advantageous when we consider corrections to the energy spectrum.

The classical expressions for the Hamiltonian of the fully interacting $\cN=1$ theory is given by the Legendre transform of (\ref{eqn:MinimalLagrangian}) according to
\be
\label{eqn:ClasicalHamiltonian}
H = \int_{T^3} d^3x \, \Tr \left\{ \frac{g^2}{2}\Pi_i\Pi^i + \frac{1}{4g^2}F_{ij}F^{ij} - \frac{i}{2}\overline{\psi} \Ga^i D_i \psi \right\} \,,
\ee
where $\Pi_i$ denotes the momentum conjugate to $A_i$. This expression contains explicit $g$-dependence which is consistent with the expectation, discussed above, that the energy spectrum should receive corrections in the interacting theory.

The $\cN=1$ supersymmetry transformation of the theory is given by
\be
\label{eqn:SupersymmetryTransformation}
\ba{lll}
\delta A_M & = & \frac{i}{2} \overline{\ep} \Ga_M \psi \\
\delta \psi & = & \frac{1}{4} F_{MN} \Ga^{MN} \ep
\ea \,,
\ee
where $\ep$ is the infinitesimal spinor parameter of the transformation. At this point it should be noted that for the $5+1$ dimensional case, where the spinor $\psi$ is not Majorana and can consequently not be identified with its conjugate $\overline{\psi}$, the transformation $\delta A_M$ must be modified to include the complex conjugate of $\overline{\ep} \Ga_M \psi$ in order for the transformations to constitute a symmetry of (\ref{eqn:MinimalLagrangian}) and for the variation of the gauge field $A_M$ to be real. For definiteness (and compactness) we will use $9+1$ notations in all computations below. The generalization to include the complex conjugate in the supersymmetry transformations is straightforward, if somewhat cumbersome, and most importantly does not influence the arguments presented in the following sections, even though the details of certain constructions need to be slightly modified.

The Noether current associated to the transformation (\ref{eqn:SupersymmetryTransformation}) is
\be
\label{eqn:Supercurrent}
J^M = \Ga^{NP} \Ga^M \Tr (\psi F_{NP} ) \,,
\ee
which is indeed conserved, $\partial_M J^M = 0$, by virtue of the equations of motion, the Pauli-Fierz identity\footnote{The Pauli-Fierz identity holds in all three dimensions $4,6$ and $10$ we consider and is the reason that supersymmetry closes in these dimensions.}  $f_{abc}\overline{\ep}\Ga_M\psi^a\overline{\psi}^b\Ga^M\psi^c = 0$, where $f_{abc}$ are the Lie algebra structure constants, and the Bianchi identity $\Ga^{MNP}\psi D_M F_{NP} = 0$. The supercharge generating (\ref{eqn:SupersymmetryTransformation}) is therefore given by
\be 
\label{eqn:Supercharge}
Q = \frac{\sqrt{\pi}}{2\sqrt{2}g^2} \int_{T^3} d^3x\,J^0\,,
\ee
where the normalization has been chosen for later convenience.

The extended supersymmetry theories in $3+1$ dimensions are recovered by splitting the vector index $M$ into the four-dimensional vector index $\mu=0,1,2,3$ and $A=4,\ldots,d$, demanding $\partial_A$ is identically zero on all fields. In the $\cN=2$ theory, the $A=4,5$ components of the gauge field are interpreted as two scalars transforming in the $\mathbf{2}$ vector representation of the $\SU(2)$ R-symmetry and the components of $\psi$ are interpreted as two Weyl spinors in $3+1$ dimensions transforming as a doublet under R-symmetry. The $\cN=4$ theory is obtained by interpreting the six transverse components of the gauge field as scalars furnishing a $\mathbf{6}$ vector representation of the R-symmetry, which in this case is $\SO(6)$. The spinor components are interpreted as four Weyl spinors (and their conjugates) transforming in the $\mathbf{4}$ ($\overline{\mathbf{4}}$) representation of the R-symmetry.

\subsection{Weak coupling expansion revisited}
Previously we considered the weak coupling expansion around an isolated flat connection $\cA = \cA_i dx^i$ in the four-dimensional theory. In terms of the higher dimensional connection one-form this expansion is simply
\be
\label{eqn:ConnectionExpansion}
A = \cA + g a \,,
\ee
where $a$ is the one-form perturbation   
\be
\label{eqn:ConnectionPerturbation}
a = a_M dx^M = a_0 dt + a_I dx^I = a_0 dt + a_i dx^i + a_A dx^A \,,
\ee
containing the perturbations to both the gauge field $A_i$ and the scalar fields $\Phi$ in the four-dimensional perspective, while the spinor field is expanded as
\be
\label{eqn:SpinorExpansion}
\psi = g \la \,,
\ee
where $\la$ is a spinor perturbation. We recall that $\cA$ is considered as a fixed background field configuration and that the dynamical fields in the weak coupling expansion description of the theory are $a$ and $\la$. As before, we denote the covariant derivative with respect to the flat connection $\cA$ by $\cD_M$ to distinguish it from the full covariant derivative $D_M$.

The Yang-Mills field strength is
\be
\label{eqn:FieldStrengthExpansion}
F_{MN} = g(\cD_M a_N - \cD_N a_M) + g^2(a_M a_N - a_N a_M)
\ee
by the flatness of $\cA$ and vanishing of its conjugate momentum, and the covariant derivative of the spinor field is
\be
\label{eqn:CovariantDerivativeSpinorExpansion}
D_M \psi = g \cD_M \la + g^2 (a_M \la - \la a_M) \,.
\ee
In terms of the perturbations $a_M$ and $\la$ the Lagrangian density may then be written as $\cL = \cL_{\rm B} + \cL_{\rm F}$, where
\be
\label{eqn:BosonicLagrangian}
\cL_{\rm B} = -\frac{1}{4g^2} \Tr \left ( F_{MN} F^{MN} \right ) = \Tr \left ( \frac{1}{2}\dot{a}_I\dot{a}^I - \frac{1}{4g^2} F_{IJ}F^{IJ} \right )
\ee
and
\be
\label{eqn:FermionicLagrangian}
\cL_{\rm F} = \frac {i}{2g^2} \Tr \left ( \overline{\psi} \Ga^M D_M \psi \right ) = \frac{i}{2} \Tr \left ( \overline{\la} \Ga^0 \dot{\la} + \overline{\la} \Ga^I D_I \la \right )
\ee
are its bosonic and fermionic parts. The conjugate momenta of the fields are $\pi^I=\dot{a}^I$ and $\pi^{(\la)}=\frac{i}{2}\overline{\la}\Ga^0$. Expressed in terms of the perturbations, the supercharge is (with the normalization chosen above) $Q=Q_0+g Q_1$ where
\be
\label{eqn:Q0PositionSpace}
Q_0 = \sqrt{\frac{\pi}{2}} \int_{T^3} d^3x\, \Tr \left( \Ga^I \la \pi_I + \Ga^J \Ga^0 \Ga^I \la \cD_I a_J \right)
\ee
and
\be
\label{eqn:Q1PositionSpace}
Q_1 = \sqrt{\frac{\pi}{8}} \int_{T^3} d^3x\, \Tr \left( \Ga^0 \Ga^{IJ} \la ( a_I a_J - a_J a_I ) \right)
\ee
and contains no higher order corrections in $g$.

Since the fields $a(x)$ and $\la(x)$ and their conjugate momenta can, in addition to time, only depend on the coordinates of the spatial torus, they may be expanded in Fourier series using the eigenfunctions (\ref{eqn:EigenvalueEquationCovariantDerivative}) forming a complete set of Lie algebra-valued functions on $T^3$. The expansion yields
\be
\label{eqn:FourierExpansionA}
a_I(x) = \sum_p a_I(p) u_p(x) \,,
\ee
\be
\label{eqn:FourierExpansionPi}
\pi_I(x) = \sum_p \pi_I(p) u_p(x) \,,
\ee
and
\be
\label{eqn:FourierExpansionLambda}
\la(x) = \sum_p \la(p) u_p(x) \,,
\ee
where the sum is over all $p$ in (\ref{eqn:SpectrumSUn}) admitted by the torus. Here, the Fourier coefficients $a_I(p)$, $\pi_I(p)$ and $\la(p)$ also carry an implicit time-dependence. The complex conjugates of the bosonic coefficients are given by
\be
\label{eqn:BosonicFourierModeComplexConjugates}
\ba{ccc}
a^I(p)^* & = & a^I(-p)\\
\pi^I(p)^* & = & \pi^I(-p)\\
\ea \,,
\ee
because of the reality of the fields $a^I$ and $\pi^I$ in position space. Both operators and states will be functions of the Fourier coefficients implying that we are working in the interaction picture, where both states and operators evolve with time. This framework typically applies to a situation like the one we are considering, where an interaction is added to the free theory. In that sense, the interaction picture is suited for the perturbation theory approach to the interacting theory we will pursue in section \ref{sec:TheInteractingTheoryAPerturbativeApproach}.

In the momentum space representation, obtained by the Fourier expansion, the components of the supercharge are given by
\be
\label{eqn:Q0MomentumSpace}
Q_0 = \sqrt{\frac{\pi}{2}} \sum_p \left( \Ga^I \la(-p) \pi_I(p) - 2 \pi i \Ga^J \Ga^0 \Ga^i p_i \la(-p) a_J(p) \right)
\ee
and
\be
\label{eqn:Q1MomentumSpace}
Q_1 = \sqrt{\frac{\pi}{8}} \sum_{p,p',p''} C_{p,p'} \delta_{p+p'+p'',0} \Ga^0 \Ga^{IJ} \la(p'') a_I(p) a_J(p')\,,
\ee 
where the orthonormality of $u_p$ and the Coulomb gauge condition, which in momentum space takes the form $p_i a^i = 0$, have been enforced. Note the appearance of the coefficients $C_{p,p'}$ related to the Lie bracket. We have retained the three-dimensional index on $p_i$ to emphasize that they are the momentum eigenvalues on $T^3$. In the higher-dimensional index notation we would equivalently have $p_I = (p_1,p_2,p_3,0,\ldots,0)$. 

From the supersymmetry algebra
\be
\label{eqn:SUSYAlgebra}
\{ \overline{\ep} Q , \overline{Q} \ep ' \} = \overline{\ep} \Ga^{\mu} \ep ' P_{\mu} \,,
\ee
where $\ep$ and $\ep '$ are bosonic spinor parameters, and the fact that $Q$ receives only linear corrections in $g$ we can conclude that the Hamiltonian of the theory is of the form $H=H_0+gH_1+g^2H_2$, with no higher order corrections than quadratic in the coupling constant. The purpose of the present paper is to consider the spectrum of the corresponding operator in the quantum theory, which leads us to the problem of quantization.

\subsection{Quantization with constraints}
Quantization of a classical gauge theory requires an assignment, consistent with any constraint present in the system, to each field in the classical theory of an operator acting on some Hilbert space $\cH$. The assignment is required to preserve the algebraic structure of the classical phase space, provided by the Poisson bracket, and may be obtained by promoting the fields to quantum operators and prescribing commutation relations according to the Dirac bracket method \cite{Dirac:1964}. In the presence of constraints causing inconsistencies that cannot be resolved by gauge fixing alone, the Dirac method modifies the canonical commutation relations to ensure consistency. The Dirac procedure corresponds to forming the symplectic quotient of the original phase space by the group of symmetries generated by the constraints, which is in fact the connected component $\mathrm{Aut}_0(P)$ of the group $\mathrm{Aut}(P)$ of bundle automorphisms, and then employing the canonical commutation relations for the quotient manifold.  

The $\cN=1$ SYM theory in $d+1$ dimensions in temporal gauge has the bosonic degrees of freedom $a^I$ subject to the additional gauge condition $\cD_I a^I = 0$. The constraints in the bosonic sector are thus $\cD_I a^I = 0$ and $\cD_I \pi^I = 0$, where the last constraint follows from the equations of motion for the gauge field. These are second class constraints, i.e.~their Poisson bracket is non-vanishing. Using the Dirac bracket then yields the commutation relation 
\be
\label{eqn:CCRPosition}
[ a^a_I(x) , \pi^b_J(x') ] =  \frac{1}{2 \pi i} \left( \de_{IJ} \delta^{ab} \de^{(3)}(x-x') - \sum_p \frac{p_I p_J}{|p|^2} u_p^b(x') \overline{u}_p^a(x) \right)
\ee
\be
\ba{ccc}
[ a^a_I(x) , a^b_J(x') ] = 0 & , & [ \pi^a_I(x) , \pi^b_J(x') ] = 0
\ea \,,
\ee
where $a,b$ are Lie algebra indices as before. The quantization in the fermionic sector, on the other hand, is not complicated by inconsistencies caused by constraints so the canonical anticommutator 
\be
\label{eqn:CARPosition}
\{ \overline{\ep} \la(x) , \overline{\la}(x') \ep ' \} = \frac{1}{\pi} \overline{\ep} \Ga^0 \ep ' \de^{(3)}(x-x')\,,
\ee
where $\ep$ and $\ep '$ are arbitrary bosonic spinors, can be used. The normalization of these (anti-)commutator relations is fixed by the fact that all the $\cN=1$ single-excitation multiplets in $d+1$ dimensions that we consider here are massless.

In the momentum space representation previously introduced, the corresponding non-vanishing (anti-)commutation relations become
\be
\label{eqn:CCRMomentum}
[ a_I(p) , \pi_J(p') ] = \frac{1}{2 \pi i} \left( \delta_{IJ} - \frac{p_I p_J}{|p|^2} \right) \delta_{p+p',0}
\ee
and
\be
\label{eqn:CARMomentum}
\{ \overline{\ep} \la(p) , \overline{\la}(p') \ep ' \} = \frac{1}{\pi} \overline{\ep} \Ga^0 \ep ' \delta_{p,p'}\,.
\ee
We will now move on to consider the quantum SYM theory on $T^3$ in the weak coupling regime, first in the limit $g \to 0$ and then in the case of small but finite $g$, using the momentum space representation and the results in the present section.

\setcounter{equation}{0}
\section{The free theory}
\label{sec:TheFreeTheory}
In this section we will consider the free SYM theory on $T^3$ at isolated flat connections, i.e.~the limit $g \to 0$, where the spectrum of the theory is known. The purpose is to describe the Hilbert space and spectrum of the theory in the higher-dimensional formalism, and introduce suitable creation and annihilation operators that will be useful when we move on to consider the interacting theory.

In general, the Hilbert space of the SYM theory is the Fock space constructed as the direct sum of tensor products of the single-excitation states of the theory, i.e.~states with transformation properties identical to those of the fundamental fields $a^I$, $\pi^I$ and $\la$. All possible such states are obtained by simply acting with the corresponding operators on the vacuum state of the theory. However, there are twice as many operators $a^I$, $\pi^I$ and $\la$ as physical degrees of freedom, since we recall that the spinor $\la$ has two real components for each physical mode, implying that half of the (linear combinations of) operators must annihilate the vacuum to yield the correct number of single-excitation states. Furthermore, the creation and annihilation operators must satisfy appropriate (anti-)commutation relations so that in our Hamiltonian formalism the corresponding states constitute eigenstates of the 4-momentum operator $P^{\mu}$, ensuring that all states in the Hilbert space have well-defined energies and momenta. Finally, the choice of particular linear combinations as creation and annihilation operators, corresponding to a choice of basis in the Hilbert space, must be compatible with the requirement from supersymmetry that the spectrum be bounded from below by zero.

\subsection{The free Hilbert space $\cH_0$}
\label{sec:TheFreeHilbertSpace}
The unique vacuum state $\vac$ of the Hilbert space $\cH_0$ of the free theory is by definition annihilated by the supercharge $Q_0$. For each momentum $p$ the second term of (\ref{eqn:Q0MomentumSpace}) suggest the introduction of the Hermitian operator
\be
\label{eqn:GammaP}
\Ga_p = |p|^{-1} p_i \Ga^0 \Ga^i\,,
\ee
acting on spinor space, which squares to unity, implying that its eigenvalues are $\pm 1$. Consequently, $\Ga_p$ induces a decomposition of $\la(p)$ according to
\be
\label{eqn:LambdaDecomposition}
\la(p) = \frac{1}{\sqrt{2 \pi}} \left( \lap(p) + \lam(p) \right)
\ee
where the components are defined by the relations
\be
\label{eqn:LambdaPlusMinus}
\Ga_p \lapm(p) = \pm \lapm(p)
\ee
or equivalently
\be
\label{eqn:LambdaPlusMinusBar}
\lapmbar(p) \Ga_p = \mp \lapmbar(p)\,.
\ee
The decomposition of $\la$ corresponds to the decomposition of the Lorentz group $\SO(d,1)$ into $\SO(1,1) \times \SO(d-1)$ in the sense that $\Ga_p$ singles out the spatial direction defined by the momentum $p_i$ from the remaining $d-1$ directions transverse to it.

Inserting the decomposition (\ref{eqn:LambdaDecomposition}) into the expression for $Q_0$ we are led to define the operators
\be
\label{eqn:AlphaAlphaDaggerDefinition}
\ba{lll}
\al_I(p) & = & \frac{1}{\sqrt{2}} |p|^{-1/2} \pi_I(p) + \sqrt{2} \pi i |p|^{1/2} a_I(p)\\
\ald_I(p) & = & \frac{1}{\sqrt{2}} |p|^{-1/2} \pi_I(p) - \sqrt{2} \pi i |p|^{1/2} a_I(p)
\ea \,,
\ee
related under complex conjugation by $\al_I(p)^* = \ald_I(-p)$, as the bosonic creation and annihilation operators. In terms of these operators the fundamental fields are given by
\be
\label{eqn:AlphaAlphaDaggerFundamentalFieldRelations}
\ba{lll}
a_I(p) & = & \frac{1}{2\sqrt{2}\pi i} |p|^{-1/2} (\al_I(p) - \ald_I(p))\\
\pi_I(p) & = & \frac{1}{\sqrt{2}} |p|^{1/2} (\al_I(p) + \ald_I(p))
\ea \,.
\ee
The operators $\al_I(p)$ and $\ald_I(p)$ obviously have the appropriate transformation properties and satisfy the commutation relation
\be
\label{eqn:CCR}
[ \al_I(p) , \ald_J(p') ] = \left( \delta_{IJ} - \frac{p_I p_J}{|p|^2} \right) \delta_{p+p',0}\,.
\ee
Thus, we may take $\ald_I(p)$ and $\al_I(-p)$ to create and annihilate a bosonic single-excitation state of momentum $+p$. It should be noted that we still let the index $I$ run over all $d$ spatial directions since there is no canonical way to eliminate the redundant gauge degree of freedom corresponding to the gauge condition $p_I a^I =0$. The extra term in the commutator compensates for this redundancy and reduces the number of independent bosonic degrees of freedom to the appropriate $d-1$.

Similarly, fermionic creation and annihilation operators can be defined using the components $\lapm(p)$. However, the details of this construction depend very much on the amount of supersymmetry, or equivalently the dimensionality $d+1$ of space-time in the higher-dimensional perspective, we consider. This is quite natural since the spinors have different properties in the cases $d=3$, $d=5$ and $d=9$. Generally, in terms of $\lapm(p)$ the expression for the supercharge takes the form
\be
\label{eqn:Q0}
Q_0 = \frac{1}{\sqrt{2}} \sum_p |p|^{1/2} \Ga^I \left( \lap(p) \al_I(-p) + \lam(-p) \ald_I(p) \right) \,.
\ee
We conclude that creation operators should be linear combinations of the components of $\lap(p)$ and annihilation operators linear combinations of the $\lam(-p)$ components, because of the structure of the expression (\ref{eqn:Q0}) in terms of the bosonic operators and the fact that $Q_0$ should annihilate the vacuum state $\vac$. 

The number of creation and annihilation operators is of course always the same as the number of physical fermionic degrees of freedom, i.e.~$d-1$. We can thus introduce the notation $\lap^m(p)$ and $\lam^m(-p)$, $m=1,\ldots,d-1$, for the operators that create and annihilate fermionic single-excitation states of momentum $+p$ and therefore satisfy the anticommutation relation
\be
\label{eqn:CAR}
\{ \lap^m(p) , \lam^n(p')\} = \de^{mn} \de_{p+p',0}\,.
\ee
In the next subsection we will consider explicitly the creation and annihilation operators in the case $\cN=4$ as an example.

According to the discussion in the introduction to this section, the full Hilbert space $\cH_0$ of the free theory is spanned by the states obtained by repeatedly acting with the creation operators $\lap^m(p)$ and $\ald_I(p)$ on the vacuum $\vac$. These states, while not orthogonal because of the form of the commutation relation (\ref{eqn:CCR}), are certainly linearly independent\footnote{Taking into account that only $d-1$ of the components of $\ald_I$ are independent.} and constitute a suitable basis $\{\mathfrak{B}_i\}$ of $\cH_0$. It should be noted that the states
\be
\label{eqn:FreeHilbertSpaceRules}
\ba{lll}
\al_I(p) \vac & = & 0\\
\lam^m(p) \vac & = & 0
\ea
\ee
equal the zero vector in the Hilbert space and are thus in fact part of $\cH_0$. Consequently, the states obtained by acting with any combination of $\al_I$, $\ald_I$, $\lap^m$ and $\lam^m$ on $\vac$ are expressible as linear combinations of the basis vectors $\mathfrak{B}_i$ using the (anti-)commutation relations. We will come back to this remark when we consider the interacting theory.    

We are now ready to consider the spatial momentum operator $P_i$ and the Hamiltonian $H_0$ of the free theory. The purpose is to verify that the creation and annihilation operators are eigenfunctions of $P_{\mu}^0=(H_0,P_i)$ and to determine the spectrum. The 4-momentum operator can be explicitly constructed using either the free supersymmetry algebra
\be
\label{eqn:FreeSUSYAlgebra}
\{ \overline{\ep} Q_0 , \overline{Q}_0 \ep ' \} = \overline{\ep} \Ga^{\mu} \ep ' P_{\mu}^0
\ee
or the corresponding classical expressions, e.g.~the expression for the Hamiltonian derived using the Legendre transformation of the Lagrangian  density. These two alternatives are equivalent and serve to determine the normalizations of creation and annihilation operators, the (anti-)commutation relations and the supercharge in (\ref{eqn:Supercharge}), by requiring that $E=|p|$ for massless single-excitation states. 

Using the commutation relations (\ref{eqn:CCR}) and (\ref{eqn:CAR}) we obtain the free Hamiltonian
\be
\label{eqn:FreeHamiltonian}
H_0 = \sum_p |p| \left( \ald_I(p)\al^I(-p) + \de^{mn}\lap^m(p)\lam^n(-p) \right)
\ee
and the spatial momentum operator
\be
\label{eqn:FreeMomentumOperator}
P_i = \sum_p p_i \left( \ald_I(p)\al^I(-p) + \de^{mn}\lap^m(p)\lam^n(-p) \right) \,.
\ee
From these expressions and the commutation relations we conclude that the states $\ald_I(p)\vac$ and $\lap^m(p)\vac$ are indeed eigenstates of both $H_0$ and $P_i$. In particular, the energy eigenvalues of the single-excitation states are $E_0 = |p|$, as required, and the complete spectrum of the theory is obtained by addition of these eigenvalues in agreement with the results in \cite{Lindman:2008}. It should be noted that even though the spectrum of the $i\cD_i$ operator, i.e.~the spectrum of allowed momenta on the torus, is non-degenerate the energy spectrum will certainly be degenerate. First, all states connected by supersymmetry transformations of course have the same energy. In addition, two distinct momenta or linear combinations of momenta in (\ref{eqn:SpectrumSUn}) may have identical absolute value, a degeneracy caused by the discrete subgroup of the Lorentz group left unbroken by our particular choice of $T^3=\RR^3 / \ZZ^3$. For an generic torus, this degeneracy is lifted since the Lorentz group is then completely broken.

\subsection{Fermionic operators in the free $\cN=4$ theory}
\label{sec:FermionicOperatorsInTheFreeN4Theory}
In this section we will consider in some detail the construction of the fermionic creation and annihilation operators of the $\cN=1$ theory in $9+1$ dimensions, which reduces to the $\cN=4$ theory in $3+1$ dimensions. In this case $\la(x)$ is a Majorana-Weyl spinor in the $\mathbf{16}$ representation of the $\SO(9,1)$ Lorentz group. The gamma matrices $\Ga^M$ satisfy the Clifford algebra $\{ \Ga^M, \Ga^N \} = 2 \eta^{MN}$ and can be chosen so that the Majorana condition implies reality of $\la(x)$, which is thus a 16 component real spinor. Furthermore, the Majorana flip in ten dimensions is
\be
\label{eqn:MajoranaFlip10D}
\overline{\psi}_1\psi_2 = \overline{\psi}_2\psi_1
\ee
for two arbitrary fermionic Majorana-Weyl spinors of opposite chirality. The equivalently identity for two Majorana-Weyl spinors of equal chirality is
\be
\label{eqn:MajoranaFlipGamma10D}
\overline{\psi}_1 \Ga^M \psi_2 = - \overline{\psi}_2 \Ga^M \psi_1\,.
\ee

Under the decomposition $\SO(9,1)\to \SO(1,1) \times \SO(8)$, corresponding to (\ref{eqn:LambdaDecomposition}), the chiral $\mathbf{16}$ spinor representation of $\SO(9,1)$ decomposes into $\mathbf{8}^+_s \oplus \mathbf{8}^-_c$. The eigenvalue under the action of $\Ga_p$ corresponds to the $\SO(1,1)$ charge of the components of $\la(p)$. Thus, the spinors $\lap(p)$ and $\lam(p)$ transform in the $\mathbf{8}_s$ and $\mathbf{8}_c$ representations of the transverse $\SO(8)_p$ respectively. Here, an index $p$ has been attached to emphasize the fact that the decomposition is characterized by the momentum $p$.

From the definition of $\Ga_p$ it is clear that the operator satisfies $\Ga_{-p} = - \Ga_p$, implying a relation between the decompositions of $\la$ for $p$ and $-p$ since $\Ga_p \lapm(-p) = \mp \lapm(-p)$. This relation simply corresponds to the fact that reversing the momentum $p$ amounts to exchanging the notions of a spinor and a cospinor while preserving the notion of spatial directions transverse to the momentum. Consequently, the spinors $\lap(-p)$ and $\lam(-p)$ transform in the $\mathbf{8}_c$ and $\mathbf{8}_s$ representations of $\SO(8)^{\perp}_p$ respectively. Because of the reality of $\la(x)$ the complex conjugates of the fermionic Fourier modes are given by
\be
\label{eqn:FermionicFourierModeComplexConjugates}
\la(p)^* = \la(-p) \,,
\ee 
which implies that the modes $\lapm$ are related through complex conjugation according to
\be
\label{eqn:ComplexConjugationLambdaPlusMinus10D}
\lapm^*(p) = \lamp(-p)\,.
\ee
The spinor bilinear identities for the Fourier modes are therefore 
\be
\label{eqn:MajoranaFlipLambdaPM10D}
\ba{lll}
\lapmbar(p) \psi & = & \overline{\psi} \lamp(-p)\\
\lapmbar(p) \Ga^M \psi & = & - \overline{\psi} \Ga^M \lamp(-p)
\ea\,,
\ee
with $\psi$ being an arbitrary fermionic Majorana-anti-Weyl and Majorana-Weyl spinor in the first and second equation respectively.

We now wish to introduce fermionic creation and annihilation operators using the components $\lap(p)$ and $\lam(p)$. As discussed above, a well-defined set of such operators must first of all have identical transformation properties which suggest $\lap(p)$ and $\lam(-p)$, since they both transform in the $\mathbf{8}_s$ representation of $\SO(8)^{\perp}_p$ and are positively charged under $\SO(1,1)_p$. The anticommutation relations for these modes, derived from (\ref{eqn:CARMomentum}), are
\be
\label{eqn:CARLambdaPM10D}
\ba{lll}
\{ \overline{\ep} \lapm(p) , \lapmbar(p') \ep ' \} & = & \overline{\ep} \Ga^0 \ep ' \delta_{p,p'}\\
\{ \overline{\ep} \lapm(p) , \lampbar(p') \ep ' \} & = & 0
\ea\,.
\ee
Using the relations (\ref{eqn:MajoranaFlipLambdaPM10D}), reducing to the physical degrees of freedom and denoting these $\lapm^m(p)$, where $m=1,\ldots,8$, we then obtain 
\be
\label{eqn:CAR10D}
\ba{lll}
\{ \lapm(p)^m , \lamp(p')^n \} & = & \de^{mn} \delta_{p+p',0}\\
\{ \lapm(p)^m , \lapm(p')^n \} & = & 0
\ea\,,
\ee
which establishes the choice of $\lap^m(p)$ and $\lam^m(-p)$ as the operators creating and annihilating a fermionic single-excitation state of momentum $+p$.

Finally, a straightforward calculation shows that $Q_0$ generates the supersymmetry transformations of the theory according to
\be
\label{eqn:Q0CommutationRelationAlpha}
\ba{lll}
[ \overline{\ep} Q_0 , \al_I(p) ] & = & -\frac{1}{\sqrt{2}}|p|^{1/2} \overline{\ep} \Ga_I\lam(p) + \frac{1}{\sqrt{2}}|p|^{-1/2}p_I \overline{\ep} \Ga^0 \lam(p) \\ 

[ \overline{\ep} Q_0 , \ald_I(p) ] & = & + \frac{1}{\sqrt{2}}|p|^{1/2} \overline{\ep} \Ga_I\lap(p) + \frac{1}{\sqrt{2}}|p|^{-1/2}p_I \overline{\ep} \Ga^0 \lap(p)
\ea
\ee
and
\be
\label{eqn:Q0CommutationRelationLambda}
\ba{lll}
\{ \overline{\ep} Q_0 , \overline{\ep}' \lam(p)\} = - \frac{1}{\sqrt{2}} |p|^{1/2} \overline{\ep} \Ga^I \Ga^0 \ep ' \al_I(p) \\
\{ \overline{\ep} Q_0 , \overline{\ep}' \lap(p)\} = - \frac{1}{\sqrt{2}} |p|^{1/2} \overline{\ep} \Ga^I \Ga^0 \ep ' \ald_I(p)
\ea \,,
\ee
which verifies that the single-excitation states indeed furnish a representation of the free supersymmetry algebra (\ref{eqn:FreeSUSYAlgebra}).

\setcounter{equation}{0}
\section{The interacting theory: A perturbative approach}
\label{sec:TheInteractingTheoryAPerturbativeApproach}
Having determined the spectrum of the theory in the free limit $g \to 0$ in the previous section, we now move on to consider the interacting theory, i.e.~where the coupling $g$ is finite but weak so that the weak coupling expansion (\ref{eqn:WeakCouplingExpansion}) is still valid. In the context of this expansion it is of course also natural to pursue a perturbative approach to the theory itself in the weak coupling regime. Generally, such an approach simply amounts to considering the operators and the states of the Hilbert space of the theory as power series in the coupling constant $g$, subject to the requirement that they reduce to the operators and states of the free theory in the limit $g \to 0$. Such power series considerations for the operators naturally implies that the corresponding eigenvalues, in particular the energy eigenvalues, are also described as power series in $g$. 

In this section we will consider the Hilbert space and the energy spectrum of the interacting theory to lowest non-trivial order in the gauge coupling. From the expression (\ref{eqn:ClasicalHamiltonian}) for the Hamiltonian we expect to find the lowest energy eigenvalue corrections at order $g^2$, which we will verify below, while the energy eigenstates receive corrections proportional to $g$. The Hamiltonian $H$ may be expressed as the free Hamiltonian $H_0$ modified by a perturbation according to  
\be
\label{eqn:HamiltonianPerturbation}
H = H_0 + g V
\ee
where the perturbation, fixed by the supersymmetry algebra as previously mentioned, is
\be
\label{eqn:SYMPerturbation}
V_{\sss SYM} = H_1 + g H_2\,,
\ee
where the parts $H_1$ and $H_2$ can be computed either using the Legendre transformation or the supersymmetry algebra as before. In what follows we will only need explicit expression for $Q_1$ and $H_1$ which are given by (\ref{eqn:Q1MomentumSpace}) and
\be
\label{eqn:H1}
H_1 = i \sum_{p,p',p''} C_{p,p'} \delta_{p+p'+p'',0} \left\{ 2 \pi p''_J a_I(p'') a^I(p) a^J(p') +\frac{1}{2} \overline{\la}(-p'')\Ga^I\la(p) a_I(p') \right\}\,,
\ee
where the fundamental fields $a_I$, $\pi_I$ and $\la$ have been used for compactness. We recall that $C_{p,p'}$ is imaginary, making $H_1$ real. The expressions for $Q_1$ and $H_1$ share two features essential for the following analysis. The first is the presence of the factor $C_{p,p'}$ which is only non-zero for distinct classes of momenta. Consequently, all terms in $Q_1$ and $H_1$ consist of mutually (anti-)commuting operators, since if $p$ and $p'$ belong to different classes of momenta $p+p'$ belongs to yet another distinct class, which implies that they have zero vacuum expectation value in the free theory $\lb 0|Q_1|0\rb = \lb 0|H_1|0\rb = 0$. The second is the cubic structure of all terms and the fact that the total momentum is zero. The importance of these properties will become apparent in the discussion to follow.

The nature of the single-excitation states is central to the analysis of the interacting theory in the present chapter and a careful consideration is therefore necessary at this point. At a first glance, the torus breaks any conformal symmetry the theory might possess in Minkowski space-time so that there is no problem of scale invariance preventing the separation of the "in" and "out" states of a scattering event. However, due to the finite size of the torus the components of the momenta take values $p_i = \ZZ + \frac{1}{2\pi}\mathrm{Arg} \, z_i$, as was previously mentioned, implying that the smallest momentum magnitudes are of the order of unity for the particular choice of torus $T^3=\RR^3/\ZZ^3$. Because of this quantization of momenta any uncertainty involved in preparing the "in" and "out" states can be assumed to be negligible compared to the separation, of order unity, between different momentum eigenvalues. The states will therefore have sharp values of the momentum and be described by plane waves in position space. Consequently, the corresponding excitations will be completely spatially delocalized and will therefore be interacting when $g$ is finite, since their wave functions certainly overlap. It is thus impossible to separate the "in" and "out" states of a scattering event allowing for them to be considered asymptotically as states of the free theory. For this reason the notion of a particle is not strictly applicable, since it implies localization in both momentum and position space, in the interacting theory and we will continue to simply use the label excitations.

The inability, due to the compactness of the spatial manifold, to separate the "in" and "out" states also has a fundamental qualitative consequence for the spectrum of the interacting theory. For a theory in Minkowski space-time (or indeed a general non-compact space-time) whose Hamiltonian can be written in the form (\ref{eqn:HamiltonianPerturbation}), as the sum of a free Hamiltonian and an interaction, we expect the full Hamiltonian $H$ to have the same spectrum as $H_0$ provided that the masses in $H_0$ are taken to be the physical masses of the theory, not the mass parameters in $H$. The reason for this expectation is the assumption that any measurement of a scattering cross-section is made at some distance from the point of interaction that is large compared to the range of the interaction, so that the states in the measurement region are effectively non-interacting. In the case of Minkowski space-time this assumption is indeed valid since the momentum spectrum is continuous, implying that the "in" and "out" states are necessarily finitely delocalized in momentum space. The excitations are therefore also localized in position space and described by wave-packets of finite extent in coordinate space. It is thus possible to measure the "in" and "out" states sufficiently far in the past or future respectively for them to be considered as non-interacting. In contrast, in the case of the torus that we are considering, we can no longer assume that measurements are made in the asymptotic region simply because it doesn't exist; the interaction region covers the entire spatial manifold. Consequently, we are in a situation where the theory described by $H$ is fundamentally different from the one described by $H_0$ since the interaction $V$ cannot be considered as simply redefining the physical masses. There is therefore no rationale for assuming that the spectrum of the interacting theory is the same as that of the free theory\footnote{It should be emphasized, however, that the appearance of new states in the Hilbert space is not expected; only a non-trivial change in the energy eigenvalues that cannot be described by simply rescaling the parameters of the theory.}.

According to the above discussion, when we consider the interacting theory we expect to find non-trivial corrections to the spectrum of the Hamiltonian in the sense that the spectrum of $H$ differs, in a non-trivial way, from that of $H_0$. In order to address this problem we now proceed to construct the interacting Hilbert space and consider the energy corrections using perturbation theory. 

\subsection{Degenerate perturbation theory}
Formally, the expressions for the energy eigenstates and eigenvalues of the interacting theory in a perturbation theory approach are given by
\be
\label{eqn:GeneralPerturbedState}
|n\rb_g = |n\rb + g |n^{\sss(1)}\rb + \cO(g^2)
\ee
and
\be
\label{eqn:GeneralPertubedEnergy}
E_n = E_n^{\sss(0)} + g E_n^{\sss(1)} + g^2 E_n^{\sss(2)} + \cO(g^3)\,,
\ee
where $|n\rb$ are the unperturbed energy eigenstates and $E_n^{\sss(0)}$ the corresponding unperturbed energy eigenvalues. In general, perturbation theory allows the calculation of eigenvalues to one order beyond that of the eigenstates. In particular, we expect to be able to compute $|n^{\sss(1)}\rb$, $E_n^{\sss(1)}$ and $E_n^{\sss(2)}$ using similar sums involving matrix elements of the unperturbed states $|n\rb$ of the free theory.

The naive way of determining the corrections to eigenstates and eigenvalues of the Hilbert space in the interacting theory is to use the results from standard perturbation theory with an arbitrary perturbation $V$, given by
\be
\label{eqn:GeneralEigenstateCorrections}
|n^{\sss(1)}\rb = \sum_{k \neq n} |k\rb \frac{\lb k|V|n\rb}{E_n^{\sss(0)}-E_k^{\sss(0)}} 
\ee
and
\bea
\label{eqn:GeneralEnergyCorrections}
E_n^{\sss(1)} & = & \lb n|V|n\rb\\
E_n^{\sss(2)} & = & \sum_{k \neq n} \frac{|\lb k|V|n\rb|^2}{E_n^{\sss(0)}-E_k^{\sss(0)}} \,.
\eea
Inserting the explicit expression for $V_{\sss SYM}$ we find that each of these terms gives two contributions, of different orders in $g$, so that the corrections in the case of the SYM perturbation take the form
\be
\label{eqn:SYMEigenstateCorrections}
|n^{\sss(1)}\rb = \sum_{k \neq n} |k\rb \frac{\lb k|H_1|n\rb}{E_n^{\sss(0)}-E_k^{\sss(0)}} 
\ee
and
\bea
\label{eqn:SYMEnergyCorrectionsLinear}
E_n^{\sss(1)} & = & \lb n|H_1|n\rb\\
\label{eqn:SYMEnergyCorrectionsQuadratic}
E_n^{\sss(2)} & = & \lb n|H_2|n\rb + \sum_{k \neq n} \frac{|\lb k|H_1|n\rb|^2}{E_n^{\sss(0)}-E_k^{\sss(0)}} \,.
\eea
  
However, in the present case we expect the above approach to encounter difficulties, since the unperturbed energy eigenvalues are degenerate and the expressions for both $|n^{\sss(1)}\rb$ and $E_n^{\sss(2)}$ therefore appear to be ill-defined. To remedy such problems one is usually required to apply degenerate perturbation theory, which amounts to choosing an appropriate basis in the subspace of degenerate unperturbed states so that all the off-diagonal matrix elements $\lb k |H_1| n \rb$ for degenerate states vanish. We must therefore consider the subspaces of degenerate states for the basis $\{\mathcal{B}_i\}$ in more detail.

In addition to the two classes of degeneracies, caused by supersymmetry and Lorentz symmetry respectively, described in the previous section there may also be accidental degeneracies, between states not related through any symmetry, in the spectrum of the free theory. In particular, such degeneracies can always be engineered by deforming the geometry of the torus. However, it is possible to show that any states $|n\rb$ and $|k\rb$ for which $H_1$ have a non-vanishing matrix element have different energies $E_n^{\sss(0)}$ and $E_k^{\sss(0)}$. The reason is that, as mentioned above, all the operators in each term of $H_1$ commute since they have momenta $p$, $p'$ and $p''=-p-p'$ belonging to distinct classes due to the presence of the $C_{p,p'}$-factor. Consequently, to obtain non-vanishing contributions to the matrix elements they must be contracted with external momenta in either $|n\rb$ or $|k\rb$ while all the remaining external momenta are contracted between $|n\rb$ and $|k\rb$. Denoting the states
\be
\ba{ccc}
|n\rb & = & |p_1,\si_1;\ldots;p_{N_n},\si_{N_n}\rb\\ 
|k\rb & = & |q_1,\tau_1;\ldots;q_{N_k},\tau_{N_k}\rb 
\ea \,,
\ee
where $p_m$, $q_m$ are the momenta and $\si_m$, $\tau_m$ denote the helicities of the corresponding creation operators, we obtain the energies as
\bea
E_n^{\sss(0)} & = & \sum^{N_n}_{m=1}|p_m|\\
E_k^{\sss(0)} & = & \sum^{N_k}_{m=1}|q_m|\,.
\eea 
A non-vanishing contribution simply implies that three of the momenta $p_m$ or $q_m$ are replaced with $p$, $p'$ and $-p-p'$ in the expressions for the energies while the rest are identified pairwise between $|n\rb$ and $|k\rb$. Thus, the energy difference between the states is given by
\be
\De E^{\sss (0)} = \pm |p| \pm |p'| \pm |p+p'|\,,
\ee
with the signs depending on the precise contraction. Because of the triangle inequality $|p+p'| < |p| + |p'|$, where we have a strict inequality since $|p|$ and $|p'|$ cannot be parallel because they belong to different classes of momenta, $\De E \neq 0$ and the energies of the states $|n\rb$ and $|k\rb$ that have a non-vanishing matrix element are necessarily different. Note that in particular this implies that the expectation value of $H_1$ in any state, not just the vacuum $\vac$, is zero, $\lb n|H_1|n \rb = 0$.

To summarize, we have showed that the matrix elements $\lb k |H_1| n \rb$ vanish for all states $|n\rb$ and $|k\rb$ that are degenerate with respect to the energy eigenvalue, a result that has two important consequences. First, the sum over states in the perturbation theory expressions can be consistently restricted to states of energies different from that of the original one, so that all potentially dangerous divergences are rendered harmless. Second, since all matrix elements in the subspace of degenerate energies are zero, the operator $H_1$ is not only diagonal but identically zero within this subspace, a property which is invariant under a change of basis. Thus, the perturbation theory expressions are valid for all choices of basis in the degenerate subspaces. The argument above is of course also valid for $Q_1$, implying that $\lb k |Q_1| n \rb = 0$ for all degenerate states, since it has a structure identical to that of $H_1$.

Before proceeding with the perturbative treatment of the interacting theory it is convenient to introduce a more compact notation for the relevant operators and states of the free Hilbert space. First of all, we can make an explicit choice of the linear combination of supercharges appearing in the supersymmetry algebra. From now on we will denote by $Q_0$ and $Q_1$ the Lorentz scalars obtained by contracting the spinor index of the components of the supercharge with an $\overline{\ep}$ parameter such that supersymmetry algebra takes the form
\bea
\label{eqn:H0SimpleNotation}
H_0 & = & Q_0^2\\
\label{eqn:H1SimpleNotation}
H_1 & = & Q_0Q_1 + Q_1Q_0\\
\label{eqn:H2SimpleNotation} 
H_2 & = & Q_1^2\,.
\eea
As was exemplified in section \ref{sec:FermionicOperatorsInTheFreeN4Theory} for the case of the $\cN=4$ theory, the action of $Q_0$ on a state in the basis $\{\mathcal{B}_i\}$ generally produces a linear combination of the states of opposite statistics. However, as we saw above, the perturbation theory results are applicable for any choice of basis so we may for simplicity choose another basis $\{|n_{\sss B,F}\rb\}$ of $\cH_0$, which is orthonormal and where the action of $Q_0$ is
\bea
\label{eqn:SuperchargeSimpleNotation}
Q_0 | n_{\sss B} \rb & = & \sqrt{E_n^{\sss(0)}} |n_{\sss F}\rb \\
Q_0 | n_{\sss F} \rb & = & \sqrt{E_n^{\sss(0)}} |n_{\sss B}\rb \,.
\eea
The labels $B$ and $F$ denote the statistics of the states and will be suppressed whenever the helicity properties of the states are not essential. This action of $Q_0$ is compatible with the orthonormality of the basis since we have
\be
\lb n_{\sss F} | n_{\sss F} \rb = \frac{1}{E_n^{\sss(0)}} \lb n_{\sss B} | Q_0^2 | n_{\sss B} \rb = \frac{1}{E_n^{\sss(0)}} \lb n_{\sss B} | H_0 | n_{\sss B} \rb = \lb n_{\sss B} | n_{\sss B} \rb \,,
\ee
so $Q_0$ preserves the normalization. Also, the orthonormality of the basis implies the completeness relation
\be
\label{eqn:BasisCompleteness}
\sum_{|n\rb \in \{|n_{\sss B,F}\rb\}} |n\rb\lb n| = 1\,.
\ee

In the following we will always use the basis $\{|n_{\sss B,F}\rb\}$ for $\cH_0$ in our considerations unless otherwise is clearly stated. Just as for the original basis $\{\mathcal{B}_i\}$ the elements of the new basis can be represented using a well-defined set of momenta and helicities corresponding to the creation operators involved, collectively denoted $\hat{n}$, according to
\be
\label{eqn:NewHilbertSpaceBasis}
|n\rb = |p_1,\si_1;\ldots;p_N,\si_N\rb = \hat{n}\vac\,.
\ee
In terms of the multi-excitation creation operators generating the $\{\mathcal{B}_i\}$ basis, $\hat{n}$ may be expressed as a linear combination of the operators with the momentum and helicity structure $\{p_1,\si_1;\ldots;p_N,\si_N\}$ for the included single-excitation creation operators $\ald_I(p)$ and $\lap^m(p)$. 

\subsection{Infinite norm states and the Stone-von Neumann theorem}
Having established that the expressions for the corrections to both eigenstates and eigenvalues of the Hamiltonian are well-defined we should, according to the standard perturbation theory prescription, finally renormalize the states $|n\rb_g$. However, considering the norm of the interacting vacuum state
\be
\label{eqn:InteractingVacuum}
\vacg = \vac - g \sum_{k\neq 0} |k\rb \frac{1}{E_k^{\sss(0)}} \lb k | H_1 | 0 \rb 
\ee
we find
\be
\label{eqn:InteractingVacuumNorm}
_g\lb 0|0 \rb_g = \lb 0|0\rb + g^2 \sum_{k\neq 0} \frac{\left| \lb k | H_1 | 0 \rb \right| ^2}{(E_k^{\sss(0)})^2} \,.
\ee
In this expression the matrix elements $\lb k |H_1| 0 \rb$ don't fall off fast enough at high momenta so that the sum diverges, which can be seen by inserting (\ref{eqn:LambdaDecomposition}) and (\ref{eqn:AlphaAlphaDaggerFundamentalFieldRelations}) into the expression (\ref{eqn:H1}) for $H_1$ and counting powers of momenta.

Consequently, the interacting vacuum is not normalizable using the norm of the Hilbert space $\cH_0$ of the free theory, and therefore does not describe a state in it. Clearly, this implies that the interacting Hilbert space $\cH_g$ cannot be identified with $\cH_0$, and that they therefore furnish unitarily inequivalent representations of the algebra (\ref{eqn:CCR}) and (\ref{eqn:CAR}) of creation and annihilation operators. This is not a situation normally considered in the context of perturbation theory in the Hamiltonian formalism, but as we will see below it is in fact a generic feature of quantum field theory.

In quantum mechanics, where the number of degrees of freedom is finite, the Stone-von Neumann theorem \cite{Rosenberg:2004} implies that there is a unique, up to unitary transformations, representation of the canonical commutation relations. Therefore, the Hilbert space of a quantum mechanical theory is unique and in particular the perturbation theory prescription always produces normalizable states that belong to that Hilbert space. The crucial assumption of the Stone-von Neumann theorem is that the Hilbert space is separable and generated by a finite number of creation operators acting on the vacuum. In the case of quantum field theory, however, the space of creation and annihilation operators is infinite dimensional and so there is in general not a unique representation, since the Stone-von Neumann theorem does not apply\footnote{Generically, the Hilbert space of a quantum field theory on a non-compact spatial manifold is also not separable. In the present case of spatial manifold $T^3$, however, the Hilbert space is in fact separable.}. Therefore, there is no reason to expect the Hilbert spaces $\cH_g$ and $\cH_{g'}$ to be identical for two different strengths $g$ and $g'$ of the coupling constant, since the corresponding representations of the operator algebra are generically unitarily inequivalent.

Fortunately, the fact that the perturbation theory produces an infinite norm vacuum state need not be the bane of the perturbative approach to the problem. We are always able to redefine the Hilbert space norm in such a way as to precisely cancel the divergence in $_g\lb 0|0 \rb_g$, a procedure which according to the above discussion implies defining a new Hilbert space. If all states obtained using (\ref{eqn:SYMEigenstateCorrections}) for the ON-basis $\{|n_{\sss B,F}\rb\}$ acquire the same divergence in their norm, they all belong to the renormalized Hilbert space and since a redefinition of the norm doesn't affect the eigenvalues, the energy corrections computed in perturbation theory are expected to give the correct energy eigenvalues in the new Hilbert space. We therefore require a more thorough understanding of the states produced by perturbation theory and the interacting Hilbert space $\cH_g$.

\subsection{Constructing the interacting Hilbert space}
In this section we consider the construction of the interacting Hilbert space $\cH_g$ to linear order in $g$ in more detail. Equivalently, such a construction amounts to a representation of the (anti-)commutation relations (\ref{eqn:CCR}) and (\ref{eqn:CAR}), which are of course independent of the coupling constant and hence valid also at finite $g$, for the interacting theory. In order to describe the interacting Hilbert space we need to specify the vacuum state and the action on this state of the creation and annihilation operators $\al_I$, $\ald_I$, $\lam^m$ and $\lap^m$. Furthermore, we need to construct a basis of linearly independent states obtained using these operators acting on $\vacg$.

Let us begin the construction by considering again the state $\vacg$ in (\ref{eqn:InteractingVacuum}), constructed using perturbation theory, to verify that it constitutes the vacuum state of the interacting Hilbert space $\cH_g$. We first note that since the free vacuum state and $H_1$ both have bosonic statistics, the sum in (\ref{eqn:InteractingVacuum}) runs over bosonic states $|k_{\sss B}\rb$ only. Since $\vacg$ reduces to $\vac$ in the limit $g \to 0$ by construction, the only properties we need to verify explicitly are that $Q=Q_0 + gQ_1$ annihilates $\vacg$ to linear order in $g$ and that the energy corrections $E_0^{\sss(1)}$ and $E_0^{\sss(2)}$ vanish so that the vacuum energy is zero to order $g^2$. In order to show this we will need to use the completeness (\ref{eqn:BasisCompleteness}) of the ON-basis $\{|n_{\sss B,F}\rb\}$.

Starting with the action of the supercharge on $\vacg$ we have
\bea
\label{eqn:SuperchargeActingOnVacuum}
Q\vacg & = & (Q_0+g Q_1)\vac - g\sum_{k \neq 0} Q_0 |k\rb \frac{1}{E_k^{\sss(0)}} \lb k | H_1 | 0 \rb + \cO(g^2)\nonumber \\
&=&gQ_1\vac -  g\sum_{k \neq 0} |k\rb \lb k | Q_1 | 0 \rb + \cO(g^2)\nonumber \\
&=&gQ_1\vac - g(1-|0\rb\lb 0|) Q_1 | 0 \rb + \cO(g^2) = 0 + \cO(g^2)\,,
\eea
where we have also used the property $\lb 0 | Q_1 | 0 \rb=0$ derived above. To verify that the energy corrections are zero we first note that $\lb 0 | H_1 | 0 \rb=0$ implies the vanishing of $E_0^{\sss(1)}$. Finally, by a computation virtually identical to (\ref{eqn:SuperchargeActingOnVacuum}), the second order correction also vanishes;
\bea
\label{eqn:QuadraticVacuumEnergyCorrection}
E_0^{\sss(2)} & = & \lb 0|H_2|0\rb - \sum_{k \neq 0} \frac{1}{E_k^{\sss(0)}} |\lb k|H_1|0\rb|^2 \nonumber \\
& = & \lb 0|Q_1Q_1|0\rb - \sum_{k \neq 0} \lb 0 | Q_1 | k \rb \lb k|Q_1|0\rb \nonumber \\
& = & \lb 0|Q_1Q_1|0\rb - \lb 0 | Q_1 (1-|0 \rb \lb 0|) Q_1|0\rb = 0 \,,
\eea
using again $\lb 0|Q_1|0\rb=0$. We can therefore conclude that $\vacg$ is indeed the vacuum state of the interacting theory.

Next, we need to consider the states $|n\rb_g$ generated by the perturbation theory formula (\ref{eqn:SYMEigenstateCorrections}), for states $|n\rb$ in the ON-basis of the free theory. Using the expression (\ref{eqn:InteractingVacuum}) for the interacting vacuum we may write a general eigenstate of the interacting theory, to first order, as
\be
\label{eqn:InteractingEigenstate}
|n\rb_g = \hat{n} \vacg + g \sum_k \hat{k} \left\{ \hat{n} \vacg \frac{\lb k | H_1| 0 \rb}{E_k^{\sss(0)}} + \vacg \frac{\lb k | H_1| n \rb}{E_n^{\sss(0)} - E_k^{\sss(0)}} \right\} + \cO(g^2) \,, 
\ee
where $\hat{n}$ and $\hat{k}$ are the linear combinations of multi-excitation creation operators corresponding to the states $|n\rb$ and $|k\rb$. In (\ref{eqn:InteractingEigenstate}) we have also used the fact that $\lb k|H_1|0\rb$ is only non-vanishing for bosonic $|k\rb$ so that $\hat{n}$ and $\hat{k}$ commute. The above expression tells us that an arbitrary state produced by perturbation theory can be expressed as a linear combination of creation operators of the free theory acting on $\vacg$. Let us consider the norm of such a multi-excitation state 
\be
\label{eqn:InteractingMultiExcitationState}
|\tilde{n}\rb_g = \lap^{m_1}(p_1)\ldots\lap^{m_f}(p_f)\ald_{I_1}(p_{f+1})\ldots\ald_{I_b}(p_{f+b})\vacg\,.
\ee
Using explicitly the expression for the (anti-)commutation relations for the creation and annihilation operators of the free theory and the expression for $\vacg$, it is straightforward to show that the norm $_g\lb \tilde{n} | \tilde{n} \rb_g$ is given by the norm $_g\lb 0\vacg$ times a finite expression involving the momenta $p_1,\ldots,p_{f+b}$ and combinatorial factors. Hence, the norm of an arbitrary state $|\tilde{n}\rb_g$ is finite in the interacting Hilbert space $\cH_g$ where the norm was rescaled so as to make $_g\lb 0\vacg$ finite. The finiteness of the norm $_g\lb \tilde{n} | \tilde{n} \rb_g$ is an expected property, considering that the only difference from the norm of the interacting vacuum is the commutation of various creation and annihilation operators, which cannot in itself add divergences. We can thus conclude that since the states $|n\rb_g$ produced by perturbation theory are expressible as linear combinations of multi-excitation creation operators acting on $\vacg$, they are all normalizable and therefore belong to $\cH_g$. Finally, the states $|n\rb$ are linearly independent and span the free Hilbert space, properties that cannot be destroyed by a continuous deformation of the theory such as a continuous change of the coupling constant like we are considering here. Consequently, the states $|n\rb_g$ are also linearly independent, since they reduce to $|n\rb$ in the limit of vanishing coupling, and constitute a suitable basis, albeit not orthonormal, of the interacting Hilbert space $\cH_g$. According to the discussion in the previous subsection this implies that the energy corrections from perturbation theory indeed give the eigenvalues of the full Hamiltonian $H=H_0+gH_1+g^2H_2$, acting in the space $\cH_g$, to quadratic order.

Before proceeding to consider these energy corrections we pause for a few remarks regarding the interacting representation of the creation and annihilation operator algebra, corresponding to the Hilbert space $\cH_g$ we constructed a basis for above. As we saw, all states of $\cH_g$ can be expressed as linear combinations of states obtained by acting repeatedly with the creation operators $\ald_I(p)$ and $\lap^m(p)$ on the interacting vacuum $\vacg$. This is precisely the situation we encountered for the free Hilbert space $\cH_0$. In order to complete the description of $\cH_g$ we also need to specify the action of the annihilation operators on $\vacg$. In the free theory, $\al_I(p)$ and $\lam^m(p)$ annihilated the vacuum state, hence their name, but this need not be the case in the interacting Hilbert space. The reason is that the only requirement on the states produced by acting with the annihilation operators on $\vacg$ is that they are expressible in terms of linear combinations of the corresponding states obtained using the creation operators, so that the number of single-excitation states match the number of physical degrees of freedom, as was discussed in the context of $\cH_0$ in section \ref{sec:TheFreeTheory}. In fact, the action of the annihilation operators on $\vacg$, to linear order in $g$, can be immediately deduced from the expression (\ref{eqn:InteractingVacuum}), yielding
\be
\label{eqn:HilbertSpaceRuleAlpha}
\al_I(p)\vacg = -g \sum_k [\al_I(p),\hat{k}] \vacg \frac{1}{E_k^{\sss(0)}} \lb k|H_1|0\rb + \cO(g^2)
\ee
and
\be
\label{eqn:HilbertSpaceRuleLamda}
\lam^m(p)\vacg = -g \sum_k [\lam^m(p),\hat{k}] \vacg \frac{1}{E_k^{\sss(0)}} \lb k|H_1|0\rb + \cO(g^2)\,,
\ee
where we have used the fact that $\vacg$ and $\vac$ are equal to zeroth order and the fact that all states $|k\rb$ for which the matrix elements are non-zero are bosonic, so that we obtain the commutator rather than the anticommutator with $\hat{k}$, regardless of the statistics of the annihilation operator. The expressions (\ref{eqn:HilbertSpaceRuleAlpha}) and (\ref{eqn:HilbertSpaceRuleLamda}) are simply new Hilbert space computation rules for evaluation of arbitrary operators acting on the states of $\cH_g$. Finally, we need to verify that the construction described above indeed produces a representation of the creation and annihilation operator algebra. However, this follows immediately from the fact that we use perturbation theory to generate the states $|n\rb_g$ as linear combinations, albeit of infinite norm in the free Hilbert space sense, of the states $|n\rb$ of $\cH_0$. Since these states certainly furnish a representation of (\ref{eqn:CCR}) and (\ref{eqn:CAR}) so will the states $|n\rb_g$. This completes our description of the interacting Hilbert space $\cH_g$.

\subsection{Energy corrections to quadratic order}
Finally, we are now in a position where we can consider the corrections of the energy eigenvalues of an arbitrary state in $\cH_0$ using the perturbation theory results (\ref{eqn:SYMEnergyCorrectionsLinear}) and (\ref{eqn:SYMEnergyCorrectionsQuadratic}). As mentioned in the introduction, we expect all energy corrections to be finite in the $\cN=1$ SYM in $9+1$ dimensions, or equivalently the maximally supersymmetric $\cN=4$ SYM in $3+1$ dimensions. The reason is that the theory is known to be finite to all orders in perturbation theory in Minkowski space, a property that is not expected to be affected by changing the spatial manifold to a torus. In the two cases with less-than-maximal supersymmetry, however, the quantum theory must be renormalized implying that the generic 1-loop contribution to the energy of a state diverges. Since the lowest order energy corrections in perturbation theory contain precisely these effects, arising from the interactions of the fields of the theory, we would generally expect them to diverge. An explicit choice of renormalization prescription is then expected to be required in order to render the corrections finite. In the present section we will consider the perturbative energy corrections to lowest non-trivial order in the coupling $g$.

Using again (\ref{eqn:SYMEnergyCorrectionsLinear}) and the fact that the expectation value of $H_1$ in any state of the free theory vanishes, we can immediately conclude that the energy correction linear in $g$ is
\be
E^{\sss(1)}_n = 0
\ee
for all states $|n\rb$, which agrees with the expectation from the expression (\ref{eqn:ClasicalHamiltonian}) for the classical Hamiltonian. Therefore, we must proceed to quadratic order to obtain the lowest energy corrections. For a state $|n_{\sss B,F}\rb$ of arbitrary statistics this is given by (\ref{eqn:SYMEnergyCorrectionsQuadratic}), which may be written as
\be
\label{eqn:E2Definition}
E^{\sss(2)}_n = \lb n_{\sss B,F} | Q_1^2 | n_{\sss B,F} \rb + \sum_{k} \frac{\left| \lb k | Q_1Q_0 + Q_0Q_1 \right| n_{\sss B,F} \rb |^2}{E_n^{\sss(0)} - E_k^{\sss(0)}}
\ee
using the expressions (\ref{eqn:H0SimpleNotation}), (\ref{eqn:H1SimpleNotation}) and (\ref{eqn:H2SimpleNotation}). Here, the sum is restricted to states $|k\rb$ with $E_k^{\sss(0)} \neq E_n^{\sss(0)}$ according to the discussion above. Using completeness of the basis $\{|n_{\sss B,F}\rb\}$ we find that the first term in (\ref{eqn:E2Definition}), containing the leading divergence, is cancelled by part of the second term yielding
\be
\label{eqn:E2}
E^{\sss(2)}_n = \sum_{k} \frac{\sqrt{E_n^{\sss(0)}}}{E_n^{\sss(0)} - E_k^{\sss(0)}} \left( \lb n_{\sss B} | H_1 | k \rb \lb k | Q_1 | n_{\sss F} \rb + \lb n_{\sss F} | H_1 | k \rb \lb k | Q_1 | n_{\sss B} \rb \right) \,,
\ee
where $|n_{\sss B}\rb$ and $|n_{\sss F}\rb$ are related by the action of $Q_0$. We note the degeneracy caused by supersymmetry is not lifted by the interaction, which is not to be expected since the perturbation $gH_1+g^2H_2$ preserves this symmetry. The same is true for the unbroken Lorentz symmetry.

Potential divergences in the energy corrections arise from the summation $\sum_{p,p',p''}$, over all internal momenta admitted by the torus, appearing in $Q_1$ and $H_1$. To have a non-vanishing contribution the two sums must effectively be the same in order to have both matrix elements $\lb n_{\sss F,B} | H_1 | k \rb$ and $\lb k | Q_1 | n_{\sss B,F} \rb$ simultaneously non-vanishing, since the states $|n_{\sss B}\rb$ and $|n_{\sss F}\rb$ have identical momentum structure. Note that the sum over states $k$ contributes no new potential divergences; it only serves to select the states $|k\rb$ corresponding non-vanishing elements for each term in the sum over internal momenta. To emphasize the perspective of the above discussion we may express the energy correction using the operator
\be
\label{eqn:EnergyCorrectionOperator}
\cO_n = \sum_{k} \frac{\sqrt{E_n^{\sss(0)}}}{E_n^{\sss(0)} - E_k^{\sss(0)}} H_1 | k \rb \lb k | Q_1 \,,
\ee
which contains the sum over internal momenta, describing the interactions of the theory causing the change in energy eigenvalues. The correction $E_n^{\sss(2)}$ is then obtained as the sum of matrix elements
\be
\label{eqn:E2EnergyCorrectionOperator}
E^{\sss(2)}_n = \lb n_{\sss B} | \cO_n | n_{\sss F} \rb + \lb n_{\sss F} | \cO_n | n_{\sss B} \rb
\ee
between external states $|n_{\sss B}\rb$ and $|n_{\sss F}\rb$.

At the moment we will primarily be concerned with the finiteness of the energy corrections and it is therefore convenient to consider the cardinality of, i.e.~the number of elements in, the intersection of the set of external momenta $\{p_1,\ldots,p_N\}$ of the state $|n\rb$ and the set of internal momenta $\{p,p',p''\}$ appearing in the operator $Q_1$, given as before by (\ref{eqn:Q1MomentumSpace}). We denote this quantity by
\be
\label{eqn:Cardinality}
\mathcal{C} = \left| \{p,p',p''\} \cap \{p_1,\ldots,p_N\} \right| \,.
\ee
Since, as we remarked in the previous paragraph, there is effectively only one summation in the expression for $E_n^{\sss (2)}$, we may take it to be the one appearing in $Q_1$ and use the cardinality $\mathcal{C}$ to characterize the terms in (\ref{eqn:E2}).

\subsubsection{$\mathcal{C} = 0$ terms}
For terms with $\mathcal{C} = 0$ all internal momenta $p$, $p'$ and $p''$ are \textit{hard}, in the sense that they don't match any of the momenta in $|n_{\sss B,F}\rb$ and are therefore taken to infinity by the sum $\sum_{p,p',p''}$. However, because all three momenta are hard all three excitations added by $Q_1$ in the factor $\lb k | Q_1 | n_{\sss B,F} \rb$ must be removed by $H_1$ in order to have $\lb n_{\sss F,B} | H_1 | k \rb$ non-vanishing. There are therefore no contractions of internal and external momenta which implies that $\mathcal{C}=0$ terms are simply proportional to the inner product of two states with opposite statistics $\lb n_{\sss F} | n_{\sss B} \rb = 0$. Consequently, all potentially diverging $\mathcal{C}=0$ terms vanish identically.

\subsubsection{$\mathcal{C} = 1$ terms}
In the case of $\mathcal{C}=1$ there is a single overlap between an internal momentum, say $p$, referred to as a \textit{soft} internal momentum, and an external momentum so that $\cO_n$ removes one of the original excitations in $|k_{\sss B,F}\rb$. Enforcing the delta function in $Q_1$ reduces the sum over internal momenta to a single sum $\sum_{p'}$, where the terms are functions of scalar products of $p'$ and $\{p_1,\ldots,p_N\}$, powers of $|p'|$ and the factor $C_{p,p'}$. The precise expressions for the individual terms may be obtained using the (anti-)commutation relations between the creation and annihilation operators of the free theory for each state $|n_{\sss B,F}\rb$. As before, the remaining hard momentum $p'$ is summed over all values admitted by the torus and the sum is expected to diverge from counting powers of $p'$. However, it is possible to show that this contribution is in fact, not only finite, but identically vanishing.

\begin{figure}[!ht]
\begin{center}
\begin{overpic}[scale=0.7]{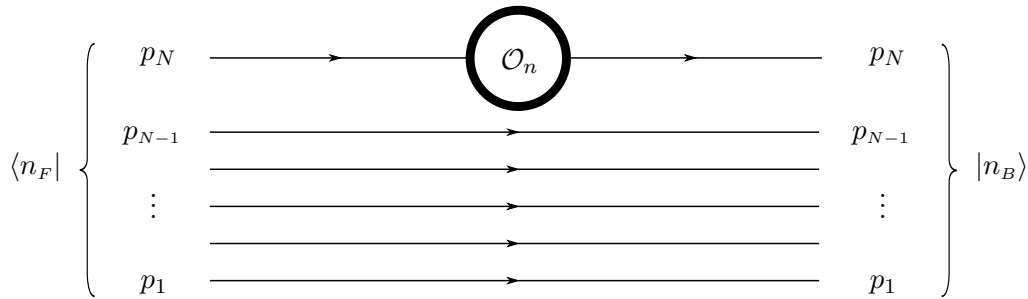}
\put(-8,14){$\lb n_{\sss F} |$}
\put(7,27){$p_N$}
\put(5,18){$p_{\sss N-1}$}
\put(8,9){$\vdots$}
\put(7,1){$p_1$}
\put(48,26){$\mathcal{O}_n$}
\put(102,14){$|n_{\sss B}\rb$}
\put(90,27){$p_N$}
\put(88,18){$p_{\sss N-1}$}
\put(91,9){$\vdots$}
\put(90,1){$p_1$}
\end{overpic}
\caption{The $\mathcal{C}=1$ contribution for arbitrary state $|n_{\sss B,F}\rb$ reduces to the single-excitation case since only one excitation participates in the interaction described by $\cO_n$.}
\label{fig:SpectatorExcitations}
\end{center}
\end{figure}

The vanishing of the $\mathcal{C}=1$ contribution can be understood as follows: Consider the terms where, say, $p=p_N$ and $\{p',p''\} \cap \{p_1,\ldots,p_{N-1}\} = \emptyset$\footnote{All $\mathcal{C}=1$ terms are of course contained in this consideration by simply relabelling the momenta.}. This implies that $p_1,\ldots,p_{N-1}$ denote the momenta of excitations that are not involved in the interaction described by $\cO_n$ but whose corresponding operators are contracted directly between the external states $\lb n_{\sss F,B}|$ and $|n_{\sss B,F}\rb$. Thus, these excitations are to be considered as spectator excitations that have no influence on the matrix elements, which implies that the $\mathcal{C}=1$ contribution to $E_n^{\sss(2)}$ for an arbitrary state $|n_{\sss B,F}\rb$ reduces to that of a single-excitation state, as illustrated in figure \ref{fig:SpectatorExcitations}.

However, the single-excitation states, being massless and therefore having four-momentum $P_{\mu} = (|p|,0,0,|p|)$ in some suitable frame, furnish a $\frac{1}{2}$-BPS representation of the free supersymmetry algebra (\ref{eqn:FreeSUSYAlgebra}). In the case of $\cN=1$ SYM theories in arbitrary dimension that we are considering here, there is no possible Higgs field, so the single-excitation states remain massless in the interacting theory and therefore continue to furnish a $\frac{1}{2}$-BPS multiplet of the full supersymmetry algebra (\ref{eqn:SUSYAlgebra}). The momentum operator $P_i$ does not receive any corrections in the interacting theory, and the single-excitation states are therefore protected from energy corrections to all orders in perturbation theory because $P_0 = |p|$. In particular, for the single-excitation states there are no terms of cardinality $\mathcal{C} > 1$ which implies that the $\mathcal{C}=0$ and $\mathcal{C}=1$ contributions must cancel to give a vanishing energy correction to quadratic order in $g$. However, as we previously argued, the contribution from $\mathcal{C}=0$ terms is zero for an arbitrary state and consequently the $\mathcal{C}=1$ contribution for a single-excitation state must also be vanishing. This, finally, allows us to conclude that the $\mathcal{C}=1$ contribution to the energy correction $E_n^{\sss(2)}$ for arbitrary state vanishes identically by virtue of the discussion above.

\subsubsection{$\mathcal{C} = 2,3$ terms}
In the remaining cases, $\mathcal{C}=2$ and $\mathcal{C}=3$, all internal momenta are in fact soft. For the $\mathcal{C}=2$ terms the reason is that the $\delta_{p+p'+p'',0}$ fixes the last remaining momentum in the sum over internal momenta. Thus, there are no remaining sums over momenta and therefore no divergences because the external momenta $\{p_1,\ldots,p_N\}$ are to be considered arbitrary but fixed. The contributions to the energy correction $E_n^{\sss(2)}$ from these $\mathcal{C}=2,3$ terms are therefore finite. Since, as we saw above, the contributions from $\mathcal{C} \leq 1$ vanish identically for all states we conclude that the energy correction $E_n^{\sss(2)}$ is finite for an arbitrary state $|n_{\sss B,F}\rb$, a statement which is of course invariant under a change of basis in the interacting Hilbert space $\cH_g$.

To conclude we note that the properties essential for the above arguments are the trilinearity of the $H_1$ and $Q_1$ operators, and the fact that the fields in each term of these two quantities are mutually commuting due to the presence of the $C_{p,p'}$ factor. These properties are not affected by the modification of the supersymmetry transformations, discussed in section \ref{sec:TheMinimallySupersymmetricPerspective}, to accommodate the $5+1$ dimensional Weyl spinors. Furthermore, the free theory serving as the starting point of the perturbation theory approach is also essentially identical (except for the number of degrees of freedom) for all SYM theories in $3+1$ dimensions, even though the explicit construction of the fermionic creation and annihilation operators differs. Consequently, the considerations described in the present section generalize to $3+1$ dimensional SYM theories with arbitrary $\cN$. 

\setcounter{equation}{0}
\section{Results and discussion}
In this paper we considered supersymmetric Yang-Mills theory with $G=\SU(n)$ in the space-time $T^3\times\RR$. Theories with extended supersymmetry were described by dimensional reduction of higher dimensional SYM theories. The complications arising from the finite size of the compact spatial manifold forced us to analyze the theory in the fully interacting region even in the weak coupling regime. Using perturbation theory we constructed the interacting Hilbert space and corrections to the energy spectrum to lowest non-trivial order. In general, such an approach is not expected to produce finite energy corrections or even a consistent description of the interacting Hilbert space. However, we saw that using the structure of the terms $H_1$ and $H_2$ imposed by supersymmetry it was possible to consistently construct the Hilbert space and show that the energy corrections are finite to $\cO(g^2)$. In fact, even though the argument depends crucially on the presence of supersymmetry, it is independent of the number $\cN$ of supersymmetry generators in the $3+1$ dimensional perspective.

An explicit computation of the finite part of the energy corrections was not attempted above. Should such a computation be undertaken it would be desirable to consider a more general geometry of the torus, $T^3=\RR^3/\La$ for some lattice $\La$, to investigate the dependence of the energy corrections on the shape of the torus. In this case the momenta would be given by the sum of a vector in the the reciprocal lattice $\La^*$ and another vector inversely proportional to the size of the torus, corresponding to the non-abelian part of $\cD_i$. In particular, in the limit where the size of the torus becomes large we expect to reach a point where the uncertainty involved in preparing a state becomes comparable to the separation between momenta forcing us to consider states of finite extent in momentum space. Consequently, in this limit we expect to be able to separate the "in" and "out" states of scattering events to a non-interaction region, at least when the energies of the states are large compared to the energy scale set by the torus, according to the discussion in section \ref{sec:TheInteractingTheoryAPerturbativeApproach}.

Finally, we could consider extending the analysis in the present paper to arbitrary simple $G$ for which the moduli space of flat connections contain isolated points suitable for the weak coupling expansion. Generically, these moduli spaces contain higher-dimensional components in addition to the isolated ones. However, as long as we restrict considerations to a perturbative analysis of the weak coupling regime these components are not relevant for the theories located at the isolated vacua. The use of perturbation theory would, however, be complicated by degeneracies in the momentum spectrum, implying that several Lie algebra generators correspond to the same momentum. In particular, the diagonalization of $H_1$ in the subspace of degenerate states of the free Hilbert space would become more involved. However, this complication corresponds to a technicality in the application of degenerate perturbation theory and should be possible to incorporate in the analysis.

\section*{Acknowledgement}
The author gratefully acknowledges M\aa ns Henningson for suggesting the problem and for invaluable guidance and feedback, Niclas Wyllard for enlightening discussions and comments, and Josef Lindman H\"ornlund for reading the manuscript. This work was supported by a grant from the G\"oran Gustafsson foundation. 


\bibliographystyle{custom_unsrt}
\bibliography{Bibliography}

\begin{thebibliography}{20}
\bibitem{Lindman:2008}
J. Lindman H\" ornlund, F. Ohlsson, {\em The weak coupling spectrum around isolated vacua in $\cN=4$ super Yang-Mills on $T^3$ with any gauge group}, JHEP {\bf 07} (2008) 077, \href{http://arxiv.org/abs/0804.0503}{\tt arXiv:0804.0503 [hep-th]}.

\bibitem{Borel:1953}
A. Borel, J.-P. Serre, {\em Sur certain sous-groupes des groupes de Lie compacts}, Comment. Math. Helv {\bf 27} (1953) 128.

\bibitem{Borel:1961}
A. Borel, {\em Sous-groupes commutatifs et torsion des groupes de Lie compacts connexes}, Tohoku Math. J. {\bf 13} (1961) 216.

\bibitem{Borel:1999}
A. Borel, R. Friedman, J.W. Morgan, {\em Almost commuting elements in compact Lie groups}, \href{http://arxiv.org/abs/math/9907007}{\tt arXiv:math/9907007 [math.GR]}.

\bibitem{Witten:1998}
E. Witten, {\em Toroidal compactification without vector structure}, JHEP {\bf 02} (1998) 006, \href{http://arxiv.org/abs/hep-th/9712028}{\tt arXiv:hep-th/9712028}.

\bibitem{Keurentjes:1998}
A. Keurentjes, A. Rosly, A. Smilga, {\em Isolated vacua in supersymmetric Yang-Mills theories}, Phys. Rev. {\bf D58} (1998) 081701, 	\href{http://arXiv.org/abs/hep-th/9805183}{\tt arXiv:hep-th/9805183}.

\bibitem{Kac:1999}
V.G. Kac, A.V. Smilga, {\em Vacuum structure in supersymmetric Yang-Mills theories with any gauge group} in M.A. Shifman (ed.), {\em The many faces of the superworld}, 185-234, World Scientific Publishing Co. Pte. Ltd., Singapore, 2000, \href{http://arXiv.org/abs/hep-th/9902029}{\tt arXiv:hep-th/9902029}.

\bibitem{Keurentjes:1999a}
A. Keurentjes, {\em Non-trivial flat connections on the 3-torus I: $G_2$ and the orthogonal groups}, JHEP {\bf 05} (1999) 001, \href{http://arXiv.org/abs/hep-th/9901154}{\tt arXiv:hep-th/9901154}.

\bibitem{Keurentjes:1999b}
A. Keurentjes, {\em Non-trivial flat connections on the 3-torus II: The exceptional groups $F_4$ and $E_{6,7,8}$}, JHEP {\bf 05} (1999) 014, \href{http://arXiv.org/abs/hep-th/9902186}{\tt arXiv:hep-th/9902186}.

\bibitem{Witten:2000}
E. Witten, {\em Supersymmetric index in four-dimensional gauge theories}, Adv. Theor. Math. Phys. {\bf 5} (2002) 841, \href{http://arxiv.org/abs/hep-th/0006010}{\tt arXiv:hep-th/0006010}.

\bibitem{Henningson:2007a}
M. Henningson, N. Wyllard, {\em Low-energy spectrum of $\cN = 4$ super-Yang-Mills on $T^3$: flat connections, bound states at threshold and S-duality}, JHEP {\bf 06} (2007) 084, \href{http://arXiv.org/abs/hep-th/0703172}{\tt arXiv:hep-th/0703172}.

\bibitem{Henningson:2007b}
M. Henningson, N. Wyllard, {\em Bound states in $\cN = 4$ SYM on $T^3$: $\Spin(2n)$ and the exceptional groups}, JHEP {\bf 07} (2007) 001, \href{http://arXiv.org/abs/0706.2803}{\tt arXiv:0706.2803 [hep-th]}.

\bibitem{Henningson:2008}
M. Henningson, N. Wyllard, {\em Zero-energy states of $\cN = 4$ SYM on $T^3$: S-duality and the mapping class group}, JHEP {\bf 04} (2008) 066, \href{http://arXiv.org/abs/0802.0660}{\tt arXiv:0802.0660 [hep-th]}.

\bibitem{Witten:1995}
E. Witten, {\em Bound states of strings and $p$-branes}, Nucl. Phys. {\bf B460} (1995) 335, \href{http://arxiv.org/abs/hep-th/9510135}{\tt 	arXiv:hep-th/9510135}.

\bibitem{Sen:1996}
A. Sen, {\em A note on marginally stable bound states in type II string theory}, Phys. Rev. {\bf D54} (1996) 2964, \href{http://arxiv.org/abs/hep-th/9510229}{\tt arXiv:hep-th/9510229}.

\bibitem{Brink:1977}
L. Brink, J.H. Schwarz, J. Scherk, {\em Supersymmetric Yang-Mills theories}, Nucl.Phys. {\bf B121} (1977) 77.

\bibitem{Dirac:1964}
P. A. M. Dirac, {\em Lectures on Quantum Mechanics}, Yeshiva University, New York, 1964.

\bibitem{Rosenberg:2004}
J. Rosenberg, {\em A Selective History of the Stone-von Neumann Theorem}, in R.S. Doran, R.V. Kadison (ed.), {\em Operator algebras, quantization, and noncommutative geometry}, 123-158, Contemp. Math. {\bf 365}, Amer. Math. Soc., Providence, 2004. 

\end{thebibliography}


\end{document}